\documentclass[useAMS,usenatbib]{mn2e}

\usepackage{times}  
\usepackage{listings}
\usepackage{graphics}
\usepackage{subfigure}
\usepackage{graphicx}
\usepackage{graphicx,xspace}
\usepackage{amssymb}
\usepackage{amsmath}
\usepackage{aas_macros}
\usepackage{lscape}
\usepackage{rotating}
\usepackage{placeins}
\usepackage{natbib}
\usepackage{color}
\usepackage{cuted}
     
\newcommand{\eagle}{{\sc{eagle}}\xspace}

\newcommand{\gadget}{{\sc{gadget-3}}\xspace}

\newcommand{\subfind}{{\sc{subfind}}\xspace}

\title[How gas flows shape the SHM relation]
{How gas flows shape the stellar-halo mass relation in the EAGLE simulation}

\author[P. D. Mitchell \& J. Schaye]{
\newauthor Peter D. Mitchell\thanks{\rm E-mail: mitchell@strw.leidenuniv.nl}$^{1}$,
Joop Schaye$^{1}$
\\
$^{1}$Leiden Observatory, Leiden University, P.O. Box 9513, 2300 RA Leiden, the Netherlands\\
}

\begin{document}
\date{\today}
\pagerange{\pageref{firstpage}--\pageref{lastpage}} \pubyear{2021}
\maketitle
\label{firstpage}


\begin{abstract}
The difference in shape between the observed galaxy stellar mass function
and the predicted dark matter halo mass function is generally explained
primarily by feedback processes. 
Feedback can shape the stellar-halo 
mass (SHM) relation by driving gas out of galaxies, by modulating the 
first-time infall of gas onto galaxies (i.e., preventative feedback), and 
by instigating fountain flows of recycled wind material. 
We present and apply a method to disentangle these effects for hydrodynamical simulations 
of galaxy formation. 
We build a model of linear coupled differential
equations that by construction reproduces the flows of gas onto and out 
of galaxies and haloes in the \eagle cosmological simulation. 
By varying 
individual terms in this model, we isolate the relative effects of star formation, 
ejection via outflow, first-time inflow 
and wind recycling on the SHM relation.
We find that for halo masses $M_{200} < 10^{12} \, \mathrm{M_\odot}$
the SHM relation is shaped primarily by a combination of ejection 
from galaxies and haloes, while
for larger $M_{200}$ preventative feedback is also important. 
The effects of
recycling and the efficiency of star formation are small.
We show that if, instead of $M_{200}$, we use the cumulative
mass of dark matter that fell in for the first time, the
evolution of the SHM relation nearly vanishes.
This suggests that the evolution is due
to the definition of halo mass rather than to an evolving physical
efficiency of galaxy formation.
Finally, we demonstrate that the mass in the circum-galactic
medium is much more sensitive to gas flows, especially recycling,
than is the case for stars and the interstellar medium.
\end{abstract}


\section{Introduction}

With modern multi-wavelength extra-galactic surveys it is now possible to infer the buildup
of galaxy stellar masses across cosmic time \cite[e.g.,][]{Bell03,Drory05,Ilbert10,Muzzin13}. 
Various empirical methods have in turn been developed to link these observations to the 
buildup of dark matter haloes, thought to host galaxies, 
using predictions from the $\Lambda$ cold dark matter ($\Lambda$CDM) cosmological model. These methods
include halo occupation distribution modelling \cite[e.g.,][]{Peacock00,Berlind02}, 
simple abundance matching \cite[e.g.,][]{Vale04,Conroy06}, and more sophisticated
forward modelling techniques \cite[e.g.,][]{Yang12,Moster18,Behroozi19}. These models
convincingly demonstrate that $\Lambda$CDM is consistent with the observed abundances
and clustering of galaxies.

\begin{figure}
\includegraphics[width=20pc]{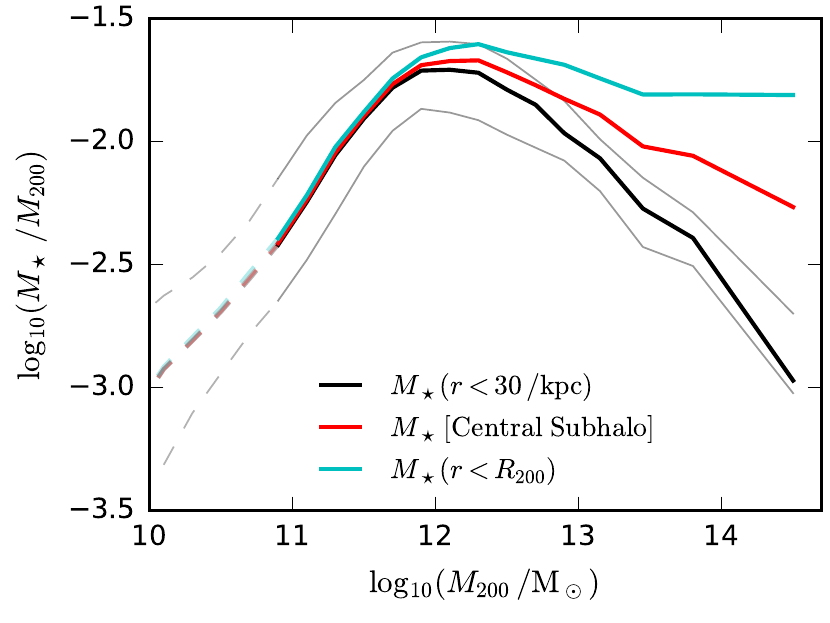}
\caption{The SHM relation, defined as the ratio of galaxy stellar mass 
to halo mass, plotted as a function of halo mass for central galaxies.
Data are taken from the Reference \eagle simulation at $z=0$ \protect \cite[][]{Schaye15}.
Grey, black and grey lines show respectively the $16$, $50$, and $84^{\mathrm{th}}$ percentiles of
the distribution, defining galaxy stellar mass as the mass within
a spherical aperture of radius $30 \, \mathrm{kpc}$.
The red line shows the median relation if the stellar mass is instead defined
as the sum of all stellar particles within the central subhalo
(this is the galaxy stellar mass definition we use in later figures).
The cyan line shows the median relation if all star particles 
within the halo virial radius ($R_{200}$) are selected.
Lighter dashed lines indicates the halo mass range where galaxies contain on average fewer than $100$ star particles.
The conversion of baryons into stars is not strongly dependent on halo mass for 
$M_{200} > 10^{12} \, \mathrm{M_\odot}$, once satellite galaxies
and the diffuse stellar halo are accounted for.
}
\label{shm_eagle}
\end{figure}

An important product of such modelling is the inferred median relationship between
galaxy stellar mass and halo mass \cite[e.g.,][]{Behroozi10,Moster10}. This
is commonly expressed as the ratio of galaxy stellar mass over halo mass 
(i.e., proportional to the efficiency
with which baryons are converted into stars) as a function of halo mass, 
which we hereafter refer to as the stellar-halo mass (SHM) relation. The ratio
of central galaxy stellar mass to halo mass is inferred to rise roughly linearly
with increasing halo mass up until the characteristic mass scale 
$\approx 10^{12} \, \mathrm{M_\odot}$, above which the ratio decreases 
\cite[e.g.,][]{Yang12,Wang13,Lu15b,RodriguezPuebla17,Kravtsov18,Moster18,Behroozi19}.
This mass dependence is shown in Fig.~\ref{shm_eagle}, using data for simulated
galaxies from the \eagle cosmological hydrodynamical simulation \cite[][]{Schaye15}.

For halo masses $<10^{12} \, \mathrm{M_\odot}$, the shape of the SHM relation is conventionally 
explained by stellar feedback;  the efficiency of feedback increases
with decreasing halo mass as it becomes easier energetically to overcome 
the gravitational potential \cite[e.g.,][]{Larson74,Dekel86,Cole91,White91}.
For halo masses $>10^{12} \, \mathrm{M_\odot}$, a reduction in the
efficiency with which baryons are converted into stars is generally
explained via the longer associated radiative cooling timescales
in massive, virialised systems \cite[][]{Rees77,White78}, combined
with the effects of efficient energy injection from an accreting central 
supermassive black hole \cite[e.g.,][]{Tabor93,Silk98,Bower06,Croton06}.
Notably, as shown by the cyan line in Fig.~\ref{shm_eagle}, the SHM ratio
actually does not decrease substantially for $M_{200} > 10^{12} \, \mathrm{M_\odot}$
once satellite galaxies and the diffuse stellar halo are included into the SHM
numerator (as they are by convention for the denominator).
It is more correct therefore to state that AGN feedback inhibits 
star formation in massive haloes, without strongly reducing the conversion
of baryons into stars compared to lower-mass haloes.

This overall picture has been fully realised and validated with various implementations
of semi-analytic galaxy formation models 
\cite[e.g.,][]{Bower06,Croton06,Somerville08}, and more recently
by hydrodynamical simulations 
\cite[e.g.,][]{Schaye10,Dubois14,Hirschmann14b,Vogelsberger14,Schaye15,Pillepich18,Dave19}.
Modern hydrodynamical simulations now convincingly reproduce the observationally inferred
SHM relation, given uncertainties in both observations and in the energy
of feedback processes that is able to mechanically power galactic outflows 
before being lost to radiation.

It has long been recognised that galaxies must continually accrete diffuse
gas from their surrounding environments in order to explain the observed
chemical abundances of stars \cite[e.g.,][]{Larson72}, and the relatively
short gas depletion timescales of star-forming galaxies \cite[e.g.,][]{Bauermeister10}.
This picture is strongly supported by cosmological hydrodynamical simulations.
Simulations also demonstrate that feedback processes could plausibly affect
diffuse gas accretion rates onto galaxies, both positively, by injecting
metals into the circum-galactic medium, CGM, and inter-galactic medium, IGM
(facilitating radiative cooling), and negatively, as galactic winds exert
thermal over-pressure and kinetic ram pressure onto the surrounding gas 
\cite[e.g.,][]{VanDeVoort11,FaucherGiguere11,Nelson15,Correa18b}. 
In addition, gas ejected from galaxies can in principle be later re-accreted,
forming a distinct galactic wind recycling contribution to galaxy growth
\cite[e.g.,][]{Oppenheimer08,Oppenheimer10,Ubler14,AnglesAlcazar17,VanDeVoort17b,Mitchell20b}.

The magnitude of these effects remains uncertain however, as does their
differential impact on galaxies as a function of halo mass. It follows
then that we do not have a full understanding of how the SHM relation
is shaped respectively by the fluxes of accreting gas that is inflowing 
for the first time, accreting gas that is recycled, and gas that is
outflowing from the ISM in a galactic wind. In semi-analytic galaxy
formation models for example, the impact of stellar feedback on first-time
inflow gas is generally neglected \cite[see ][for a discussion]{Pandya20}, such that
the shape of the SHM relation is set by the dependence of the
galaxy-scale mass loading factor (defined as the galactic outflow rate divided
by the star formation rate) on halo mass for $M_{200} < 10^{12} \, \mathrm{M_\odot}$,
and by the impact of AGN suppressing cooling in more massive haloes
\cite[][]{Mitchell16}.

For hydrodynamical simulations it is not trivial to infer the
relative importance of these effects. As an example,
in \cite{Mitchell20a} we show gas outflow rates are (surprisingly) in general
larger at the halo virial radius than at the ISM-CGM interface,
and that the outflow rates at the two scales have qualitatively
different dependencies on halo mass, and on redshift \cite[see also][]{Pandya21}. Outflows
at both scales presumably affect the SHM relation, either
by removing gas from the galaxy, or by indirectly preventing
gas from reaching it, but it is not obvious what
the relevant importance of these effects is.
While one can vary the parameters of the feedback models
implemented in a simulation and try to interpret the resulting changes
in galaxy properties \cite[e.g.,][]{Correa18b}, any changes will
affect both outflowing and inflowing gas (across a range of scales) at the same time. This obfuscates
the interpretation of the relative importance of first-time
accretion, recycled accretion, and outflows at different spatial scales.

In this paper, we introduce a method to investigate how gaseous
inflows and outflows shape the SHM relation. We make
use of a complete set of measurements of galactic inflow and outflow
rates \cite[][]{Mitchell20b,Mitchell20a} from the \eagle 
hydrodynamical simulation project \cite[][]{Schaye15}.
Following \cite{Neistein12}, we then build a model 
of mass conservation equations, where each model term is
set to the corresponding average value measured from \eagle.
The structure of the model mimics that
of conventional semi-analytic galaxy formation models, in that it
tracks the integrated baryonic mass in different discrete components
(i.e., CGM, ISM, stars), following the underlying  
hierarchical assembly of dark matter haloes and subhaloes.
By varying the various terms in the model, 
we are able isolate how
different parts of the network of gas flows around galaxies
shape the SHM relation in the \eagle simulation. 

Since \eagle is itself a (more complex) model, our conclusions derived
from this approach will accordingly be model dependent. Nonetheless,
this study still provides a physically plausible picture 
(in terms of feedback energetics, momentum input, etc)
for the connection of the SHM relation to
gaseous inflows and outflows. The model framework could be
extended to consider measurements from other cosmological simulations, or 
constrained directly against observations using statistical inference,
using simulation results as a prior \cite[as in][for example]{Mitra15}.

The layout of this paper is as follows. We describe the \eagle simulations,
the measurements of inflow and outflow rates, and our modelling 
methodology in Section~\ref{method_sec}. We outline a simplified picture
for the factors shaping the SHM relation in Section~\ref{dave_sec}.
Our main analysis of the SHM relation is presented in Section~\ref{results_sec}.
A short extension of the analysis to the CGM and ISM is presented
in Section~\ref{ism_cgm_sec}, and we summarise our results in
Section~\ref{summary_sec}.

\section{Methods}
\label{method_sec}

\subsection{Simulations}
\label{simul_sec}

Our analysis is based on the \eagle project \cite[][]{Schaye15,Crain15}, which has been publicly 
released \cite[][]{McAlpine16}, and which includes a suite of cosmological simulations of 
various volumes, resolutions, and model variations. \eagle uses a modified version
of the \gadget code \cite[last described in][]{Springel05b} to simulate cubic, periodic
regions of the Universe, solving for the equations of hydrodynamics and gravity,
using smoothed particle hydrodynamics (SPH). 
A $\Lambda$CDM cosmological model is assumed, with parameters taken from \cite{Planck14}. 
Simple ``subgrid'' models are included to account for the effect of 
physical processes that are not resolved 
and/or otherwise explicitly simulated. These include star formation, stellar evolution and feedback, 
supermassive black hole (SMBH) seeding, dynamics and growth, feedback from active galactic 
nuclei (AGN), and radiative cooling and heating.

Full details of each aspect of this modelling are given by \cite{Schaye15}
and references therein. A salient aspect is that both stellar and AGN feedback are modelled
as thermal energy injection by a fixed temperature difference, with the temperature
difference set high enough that spurious radiative cooling is mitigated given the
unresolved nature of the simulated ISM \cite[][]{Booth09,DallaVecchia12}. This in turn
drives powerful galactic outflows that entrain gas in the circum-galactic
medium and generally lead to high outflow rates at the halo virial radius \cite[][]{Mitchell20a},
and that can significantly reduce rates of first-time gas infall \cite[][]{Mitchell20b,Wright20}.

The parameters of the Reference \eagle model are calibrated such that the simulation 
is consistent with observed present-day star formation thresholds and $\mathrm{kpc}$-scale efficiencies,
such that the simulation broadly reproduces the observed galaxy stellar mass function 
and trends of galaxy size with stellar mass, 
and likewise that supermassive black hole masses are consistent with observations at a given stellar mass. We
use the Reference \eagle model for this paper, using measurements from the largest
simulation with $1504^3$ particles, a volume of $(100 \, \mathrm{Mpc})^3$, 
dark matter particle mass of $9.7 \times 10^6 \, \mathrm{M_\odot}$, initial gas
particle mass of $1.8 \times 10^6 \, \mathrm{M_\odot}$, and maximum physical
gravitational softening of $0.7 \, \mathrm{kpc}$.

\subsection{Measurements}
\label{measure_sec}

We make use of a comprehensive set of measurements tracking gas fluxes in \eagle, as
presented in detail by \cite{Mitchell20b,Mitchell20a}. In brief, we
measure inflow and outflow rates at two scales. First, we measure fluxes at the
halo virial radius (which we will refer to as the ``halo scale'') according to which particles enter or leave
the virial radius between two consecutive stored simulation outputs.
We use $200$ simulation outputs in total, with a spacing of $\approx 120 \, \mathrm{Myr}$
at low redshift, and with the spacing becoming finer with increasing redshift \cite[see appendix A of ][]{Mitchell20a}.
We define the virial radius as the radius enclosing a mean overdensity that is $200$ times
the critical density of the Universe. We further use the \subfind algorithm
\cite[][]{Springel01,Dolag09} to define membership of particles within $R_{200}$ to 
different subhaloes, distinguishing the central
subhalo from satellite subhaloes on the basis of which subhalo contains the
particle with the lowest value of the gravitational potential. Any particles
that are within $R_{200}$ but which are not considered bound by \subfind to any
subhalo are redefined as belonging to the central subhalo.

We also measure fluxes at the boundary of the ISM (referred to as ``galaxy scale''), according
to which particles enter/leave the ISM between two consecutive outputs. 
The ISM is defined as particles that either pass the \eagle star formation
threshold based on density, temperature and metallicity 
\cite[which captures the transition from the warm, atomic to the cold, molecular gas phase:][]{Schaye04},
or that otherwise
have total hydrogen number density $n_{\mathrm{H}} > 0.01 \, \mathrm{cm^{-3}}$ and are within 
$0.5 \, \mathrm{dex}$ of the temperature floor corresponding
to the equation of state imposed on the unresolved ISM \cite[][]{Schaye08,Mitchell20a}.
In practice, with this definition most of the ISM is classified as star forming, and
the latter selection (cool gas with $n_{\mathrm{H}} > 0.01 \, \mathrm{cm^{-3}}$)
acts only to add some low-metallicity gas, roughly mimicking a selection on atomic neutral hydrogen,
and is most important in low-mass galaxies \cite[$M_{200} < 10^{11} \, \mathrm{M_\odot}$, see appendix A3 in][]{Mitchell20a}.
In addition, we measure rates of star formation and stellar mass loss
\cite[due to stellar evolution, ][]{Wiersma09}, again defined as the change in mass between
two consecutive simulation outputs.

Finally, we keep track of which gas particles were ejected from the ISM
of a galaxy, and which gas particles were ejected from a subhalo.
This information is then propagated through subhalo merger trees
(accounting for mergers) and used to determine if gas accreted
onto haloes or galaxies is recycled after having been ejected in the past.
By then aggregating all these different measurements, we
construct a complete description of baryonic assembly associated
with dark matter haloes, expressed in terms of the total mass of the
CGM (out to $R_{200}$), the ISM, stars, and any gas that has
been ejected outwards beyond $R_{200}$. For simplicity, we
define galaxy stellar masses as the sum of all stars within
a given subhalo, without applying any spatial aperture selection.

\subsection{The N12 model}
\label{n12_desc_sec}

Our objective is to understand the connection between the SHM
relation and the underlying network of gaseous inflows, outflows, star formation, and wind recycling.
One way to achieve this would involve first running multiple cosmological simulations with different
choices for subgrid models and parameters, and then assessing the impact of these choices on
gas flows and on stellar mass assembly. An alternative (and complementary) 
methodology that we introduce here is to first
measure the average behaviour in a single
cosmological simulation, and second, to then construct a model that reproduces this average
behaviour, and third, to apply variations to the model in order to understand the role and relative
importance of its various components. 

Note that our approach is not equivalent to running multiple hydrodynamical simulations with 
different choices of subgrid model parameters. With our methodology, we can modify
parts of our model to understand the isolated effect of its different components.
These modifications are (in general) not expected to mimic the effect of changing
subgrid parameters, as the model will
not capture the non-linear complexity of full cosmological simulations, and changes
in subgrid parameters typically lead to changes in multiple types of gas flows. 
Rather our methodology provides a way to modify one aspect of the parent simulation,
while holding all other aspects constant, and should therefore only be viewed as a way to understand the behaviour
of a single parent hydrodynamical simulation with one unique set of subgrid parameters (in this case, the reference \eagle simulation).

\setcitestyle{notesep={ }}
To implement this, we use the methodology introduced
by \cite{Neistein12}, and we refer hereafter to our implementation of this framework as the ``N12
model'' \cite[note that by this we are always referring to our implementation, and not the original
implementation as presented in][]{Neistein12}.
\cite{Neistein12} demonstrated that by measuring the average inflow, star formation, and outflow rates
of gas of galaxies in a cosmological simulation as a function of halo mass and redshift, it is possible
(using only this averaged information plus the halo merger tree of each galaxy) to accurately reproduce
the stellar mass of individual galaxies from the original simulation. They applied
this framework to a simulation from the OverWhelmingly Large Simulations \cite[OWLS,][]{Schaye10}.
Here we apply, with a number of modifications, the same methodology to the $(100 \, \mathrm{Mpc})^3$-volume 
simulation from the \eagle simulation project. 
Some of the most important modifications are that we split gas accretion
  between first	infall versus recycled accretion, and that we explicitly track
  outflows at the virial radius	(as well as from the ISM). In practice this makes
  our version of the N12 model more complex than the	original implementation	
  from \cite{Neistein12}. Our motivation for adding this complexity is not
  to achieve a better match to the underlying simulation \cite[][already demonstrate that adding to or reducing the complexity of the model
    does not seem to significantly affect its accuracy]{Neistein12}, but
  rather because our analysis in \cite{Mitchell20a} highlights the likely substantial importance
  of halo-scale	outflows at $R_{200}$ in \eagle, which can in some cases be an order of magnitude larger
  than at the ISM/CGM interface. Separating outflows at	the two	scales can be rationalised
  as a physical separation between the mass and energy contents of galaxy-scale outflows, where the latter
  will be ultimately responsible for setting the larger-scale mass outflow rate
  at $R_{200}$.

\setcitestyle{notesep={, }}

The basic structure of the N12 model resembles that of a typical semi-analytic galaxy formation model
\cite[e.g.,][]{Guo11,Somerville12,Lacey16,Lagos18}. The baryonic content of each dark matter subhalo in a merger tree
is split between the following components: stars ($M_\star$), the ISM ($M_{\mathrm{ISM}}$), 
the CGM out to $R_{200}$ ($M_{\mathrm{CGM}}$), and (newly in our
implementation) a reservoir of gas that has been ejected beyond (and still resides beyond) $R_{200}$ ($M_{\mathrm{ej}}^{\mathrm{halo}}$). The model
tracks the mass in each of these components, and computes the various exchanges
of mass between them. When two subhaloes in a merger tree ``merge''\footnote{This 
refers to the moment at which a satellite subhalo is sufficiently disrupted (by tidal effects)
that it can no longer be identified by \subfind, which may or may not correspond to the exact moment of a 
particular definition of a galaxy-galaxy merger. Satellite subhaloes that vanish but then reappear at a later snapshot
are not merged.}, the masses of each baryonic component are simply summed.

The mass exchanges that take place between the baryonic components are given by the following
set of ordinary differential equations:

\begingroup
\begin{multline}
\renewcommand{\arraystretch}{2.}
\begin{bmatrix}
  \dot{M}_{\mathrm{CGM}} \\
  \dot{M}_{\mathrm{ISM}} \\
  \dot{M}_{\mathrm{ej}}^{\mathrm{halo}} \\
  \dot{M}_{\star} \\
\end{bmatrix}
=
\renewcommand{\arraystretch}{2.}
\begin{bmatrix}
  f_{\mathrm{acc}}^{\mathrm{halo}} \, \frac{\Omega_{\mathrm{b}}}{\Omega_{\mathrm{m}} - \Omega_{\mathrm{b}}} \dot{M}_{\mathrm{DM}}^{\mathrm{1st}} \, -\frac{G_{\mathrm{ret}}^{\mathrm{gal}} }{t} M_{\mathrm{ej}}^{\mathrm{gal}}\\
  \frac{G_{\mathrm{ret}}^{\mathrm{gal}} }{t} M_{\mathrm{ej}}^{\mathrm{gal}}\\
  0\\
  0\\
\end{bmatrix}
+\\
\renewcommand{\arraystretch}{2.3}
\begin{bmatrix}
  -F_{\mathrm{CGM}}^{\mathrm{pr}} \frac{\strut G_{\mathrm{acc}}^{\mathrm{gal}} }{\strut t} & (\eta^{\mathrm{gal}}-\eta^{\mathrm{halo}}) \frac{\strut G_{\mathrm{SF}}}{\strut t} & \frac{\strut G_{\mathrm{ret}}^{\mathrm{halo}}}{\strut t} & 0 \\
  F_{\mathrm{CGM}}^{\mathrm{pr}} \frac{\strut G_{\mathrm{acc}}^{\mathrm{gal}} }{\strut t} & -(1-R+\eta^{\mathrm{gal}}) \frac{\strut G_{\mathrm{SF}}}{\strut t} & 0 & 0  \\
  0 & \eta^{\mathrm{halo}} \frac{\strut G_{\mathrm{SF}}}{\strut t} & -\frac{\strut G_{\mathrm{ret}}^{\mathrm{halo}}}{\strut t} & 0 \\
  0 & (1-R) \frac{\strut G_{\mathrm{SF}}}{\strut t} & 0 & 0 \\
\end{bmatrix}
\renewcommand{\arraystretch}{1.8}
\begin{bmatrix}
  M_{\mathrm{CGM}} \\[1.9ex]
  M_{\mathrm{ISM}} \\[1.9ex]
  M_{\mathrm{ej}}^{\mathrm{halo}} \\[1.9ex]
  M_{\mathrm{\star}} \\[1.9ex]
\end{bmatrix},
\label{ODE_MATRIX}
\end{multline}
\endgroup

\noindent where $M_{\mathrm{CGM}}$, $M_{\mathrm{ISM}}$, $M_{\mathrm{ej}}^{\mathrm{halo}}$ and $M_\star$ are 
respectively the masses in the CGM, ISM, the reservoir of gas ejected from the halo, and in stars. In addition,
we also track the mass of gas that has been ejected from the ISM of progenitor galaxies in
a galactic wind, $M_{\mathrm{ej}}^{\mathrm{gal}}$. $M_{\mathrm{ej}}^{\mathrm{gal}}$ is comprised of gas that is
part of the CGM, or that is part of the ejected gas reservoir located beyond the virial radius, and so
is not mutually exclusive from $M_{\mathrm{CGM}}$ and $M_{\mathrm{ej}}^{\mathrm{halo}}$.
$M_{\mathrm{ej}}^{\mathrm{gal}}$ is itself governed by the equation:

\begin{equation}
\dot{M}_{\mathrm{ej}}^{\mathrm{gal}} = \eta^{\mathrm{gal}} \frac{G_{\mathrm{SF}}}{t} M_{\mathrm{ISM}} - \frac{G_{\mathrm{ret}}^{\mathrm{gal}}}{t} M_{\mathrm{ej}}^{\mathrm{gal}}.
\label{mwind_eq}
\end{equation}

\noindent $\dot{M}_{\mathrm{DM}}^{\mathrm{1st}}$ in Eqn.~\ref{ODE_MATRIX}
is the ``smooth'' accretion rate of dark matter particles that are entering the halo for the
first time. We define ``smooth'' accretion as any particles that are accreted through $R_{200}$
while not considered bound to any other subhalo with a mass greater than $9.7 \times 10^8 \, \mathrm{M_\odot}$,
which corresponds to the mass of $100$ dark matter particles at fiducial \eagle resolution \cite[see
discussion in section 2.3 of][]{Mitchell20b}. 
$\Omega_{\mathrm{m}}$ is the cosmic matter density, and $\Omega_{\mathrm{b}}$ is the cosmic baryonic matter density.
$t$ in both Eqns~\ref{ODE_MATRIX},\ref{mwind_eq} is the age of the Universe at redshift $z$.
$F_{\mathrm{CGM}}^{\mathrm{pr}}$ is the mass fraction of the CGM 
that has \emph{not} been ejected from the ISM of a progenitor galaxy of the current subhalo. 
If we split $M_{\mathrm{ej}}^{\mathrm{gal}}$ between 
gas that belongs to the CGM, and gas that belongs to the ejected gas reservoir beyond $R_{200}$,
such that
$M_{\mathrm{ej}}^{\mathrm{gal}} = M_{\mathrm{ej}}^{\mathrm{gal}}(r<R_{200}) + M_{\mathrm{ej}}^{\mathrm{gal}}(r>R_{200})$,
then $F_{\mathrm{CGM}}^{\mathrm{pr}}$ is defined as 
$F_{\mathrm{CGM}}^{\mathrm{pr}} \equiv \frac{M_{\mathrm{CGM}}-M_{\mathrm{ej}}^{\mathrm{gal}}(r<R_{200})}{M_{\mathrm{CGM}}}$.
The remaining terms in Eqn.~\ref{ODE_MATRIX}
are then coefficients that represent the efficiency of various physical processes, such as the
efficiency of first-time gas accretion at the virial radius ($f_{\mathrm{acc}}^{\mathrm{halo}}$), the efficiency of
galactic outflows ($\eta^{\mathrm{gal}}$), and so on. The meaning of these various terms is
as follows:

\begin{itemize}

\item \textbf{First-time gas accretion at $R_{200}$ ($f_{\mathrm{acc}}^{\mathrm{halo}}$):}
the source term for baryonic accretion is $\dot{M}_{\mathrm{CGM}} = f_{\mathrm{acc}}^{\mathrm{halo}} \, \frac{\Omega_{\mathrm{b}}}{\Omega_{\mathrm{m}} - \Omega_{\mathrm{b}}} \dot{M}_{\mathrm{DM}}^{\mathrm{1st}}$,
which represents the smooth accretion of gas that has not been ejected from a progenitor subhalo in
the past. We assume that gas accretion traces dark matter accretion at this radius since, 
in the absence of strong feedback effects, we expect that $R_{200}$ approximately marks
the spatial scale where a virial shock (if present) develops, and correspondingly 
marks the location where thermal pressure gradients are expected to significantly decouple the 
further inwards accretion of gas relative to that of collisionless dark matter. 
We choose to
express this as a function of first-time smooth dark matter accretion (rather than the more
conventional choice of simply the total dark matter accretion rate) because a significant
fraction of accreting dark matter particles at the virial radius have already crossed
this threshold in the past \cite[e.g.,][]{Wright20} and are being re-accreted, at which point 
the dynamics should be already decoupled from that of the gas. 
$f_{\mathrm{acc}}^{\mathrm{halo}}$ is then a coefficient that represents
any deviation from the case where first-time gas accretion traces that of dark matter. In \cite{Mitchell20b},
we show that $f_{\mathrm{acc}}^{\mathrm{halo}}$ drops below unity in low-mass haloes ($M_{200} < 10^{11} \, \mathrm{M_\odot}$),
and \cite{Wright20} demonstrate explicitly that this is because of feedback processes in the
\eagle simulations \cite[i.e. ``preventative'' feedback; see also][]{VanDeVoort11}. In higher-mass haloes, $f_{\mathrm{acc}}^{\mathrm{halo}}$ 
can actually exceed unity as gas that is prevented from being accreted onto low-mass progenitors
at early times can catch up to cumulative dark matter accretion by being accreted onto 
more massive descendant haloes later on \cite[][]{Mitchell20b}.

\vspace{0.3cm}

\item \textbf{First-time galaxy-scale accretion ($G_{\mathrm{acc}}^{\mathrm{gal}}$):}
gas in the CGM (that has not been in the ISM in the past) is then assumed to infall onto the ISM at the rate
$\dot{M}_{\mathrm{in}} = \dot{M}_{\mathrm{ISM}} = F_{\mathrm{CGM}}^{\mathrm{pr}} G_{\mathrm{acc}}^{\mathrm{gal}} M_{\mathrm{CGM}} / t$,
where $F_{\mathrm{CGM}}^{\mathrm{pr}} M_{\mathrm{CGM}}$ is the
mass in the CGM that has not been present in the ISM of a progenitor
galaxy of the current subhalo in the past, $G_{\mathrm{acc}}^{\mathrm{gal}}$
is the dimensionless efficiency of first-time gaseous infall from
the CGM to the ISM, and $t$ is the age of the Universe at a given redshift.
$G_{\mathrm{acc}}^{\mathrm{gal}}$ is related to the characteristic timescale
for the CGM to be depleted by gas accretion onto the ISM ($\tau_{\mathrm{acc}}^{\mathrm{gal}}$)
by $G_{\mathrm{acc}}^{\mathrm{gal}} \equiv t / \tau_{\mathrm{acc}}^{\mathrm{gal}}$.
We choose to express the model coefficients related to timescales 
($G_{\mathrm{acc}}^{\mathrm{gal}}, G_{\mathrm{SF}}, G_{\mathrm{ret}}^{\mathrm{gal}}, G_{\mathrm{ret}}^{\mathrm{halo}}$) 
in this way to ensure first
that they are dimensionless (factoring out the zeroth order time dependence
as galactic and halo-scale timescales generally scale linearly with the 
age of the Universe), and second to ensure that higher values of $G$
intuitively imply higher efficiencies of gas accretion, star formation, or wind recycling.

\vspace{0.3cm}

\item \textbf{Star formation ($G_{\mathrm{SF}}$):}
gas within the ISM of galaxies is converted into stars at the rate
$\dot{M}_\star = G_{\mathrm{SF}} \, M_{\mathrm{ISM}} \, / t$.
In this case $G_{\mathrm{SF}} = t \, / \tau_{\mathrm{SF}}$,
where $\tau_{\mathrm{SF}} \equiv M_{\mathrm{ISM}} \, / \dot{M}_\star$ 
is the conventional characteristic gas depletion
time for the ISM to be depleted by star formation. 

\vspace{0.3cm}

\item \textbf{Stellar recycling ($R$):}
stars return mass to the ISM at a rate given by $\dot{M}_{\mathrm{ISM}} = R \, G_{\mathrm{SF}} \, M_{\mathrm{ISM}} / t$,
where $R$ is the recycled fraction. $R$ should depend on the full star
formation and chemical enrichment history of the galaxy, but (for
reasons of numerical convenience) we make the assumption that
$R$ can be parametrised as a function of the current
star formation rate ($G_{\mathrm{SF}} \, M_{\mathrm{ISM}} / t$).

\vspace{0.3cm}

\item \textbf{Galaxy- and halo-scale outflows ($\eta^{\mathrm{gal}}$ and $\eta^{\mathrm{halo}}$):}
gas is ejected from the ISM in galactic winds with a mass outflow rate
$\dot{M}_{\mathrm{out}} = -\dot{M}_{\mathrm{ISM}} = -\eta^{\mathrm{gal}} \, G_{\mathrm{SF}} \, M_{\mathrm{ISM}} / t$,
where $\eta^{\mathrm{gal}}$ is the galaxy-scale dimensionless mass loading factor,
and $G_{\mathrm{SF}} \, M_{\mathrm{ISM}} / t$ is the star formation
rate ($\dot{M}_\star$). Similarly, gas is ejected from the CGM to the ejected gas reservoir outside
$R_{200}$ with a rate $\dot{M}_{\mathrm{CGM}} = -\eta^{\mathrm{halo}} \, G_{\mathrm{SF}} \, M_{\mathrm{ISM}} / t$,
where $\eta^{\mathrm{halo}}$ is the halo-scale dimensionless mass loading factor.

\vspace{0.3cm}

\item \textbf{Galaxy- and halo-scale gas recycling ($G_{\mathrm{ret}}^{\mathrm{gal}}$ and $G_{\mathrm{ret}}^{\mathrm{halo}}$):}
gas that has been ejected from the ISM is assumed to return at the rate
$\dot{M}_{\mathrm{ISM}} = G_{\mathrm{ret}}^{\mathrm{gal}} \, M_{\mathrm{ej}}^{\mathrm{gal}} / t$, where
$G_{\mathrm{ret}}^{\mathrm{gal}}$ is the dimensionless efficiency of galaxy-scale
wind recycling. Similarly, gas that has been ejected from
the CGM to outside $R_{200}$ is assumed to return to the CGM at the
rate $\dot{M}_{\mathrm{CGM}} = G_{\mathrm{ret}}^{\mathrm{halo}} \, M_{\mathrm{ej}}^{\mathrm{halo}} / t$,
where $G_{\mathrm{ret}}^{\mathrm{halo}}$ is the dimensionless efficiency of halo-scale gas recycling.

\end{itemize}

Fiducial values for for these various terms are computed for individual \eagle galaxies
as described in Section~\ref{measure_sec}. Following \cite{Neistein12}, we then
then compute averages for each quantity as a function of halo mass and redshift.
Within a given mass and redshift bin, we compute the weighted mean of the associated
numerator and denominator of each term. Taking for example the galaxy-scale
outflow efficiency, this is computed as 

\begin{equation}
\langle \eta^{\mathrm{gal}} \rangle(M_{200},z) = \frac{ \sum_i^N w_i \dot{M}_{\mathrm{out},i} }{ \sum_i^N w_i \dot{M}_{\mathrm{\star},i} },
\end{equation}

\noindent where $\dot{M}_{\mathrm{out}}$ is the outflow rate and $\dot{M}_\star$ is
the star formation rate, summing over the $N$ galaxies that are present in a given bin. 
As a second example, the $G_{\mathrm{acc}}^{\mathrm{gal}}$ term is computed as

\begin{equation}
\langle G_{\mathrm{acc}}^{\mathrm{gal}} \rangle(M_{200},z) = \frac{ \sum_i^N w_i \dot{M}_{\mathrm{in},i} }{ \sum_i^N w_i F_{\mathrm{CGM}}^{\mathrm{pr}} M_{\mathrm{\mathrm{CGM}},i} / t(z) },
\end{equation}

\noindent where $\dot{M}_{\mathrm{in},i}$ is the inflow rate onto the ISM (here only including gas
that has not been in the ISM before), and $t(z)$ is the age of the Universe at a given redshift.

Each redshift bin contains $\approx 25$ simulation snapshots, and so the same galaxy
can appear multiple times inside the same bin. As such, and given that we
are computing separate means for the numerator and denominator,
we are implicitly averaging (in this case) outflow rates and star formation rates 
over the full redshift interval encompassed by each bin, smoothing out the
phase differences between star formation and outflow events that exist
within a given bin. Using mean statistics is preferable 
for low-mass galaxies that at our resolution have a mix of zero and non-zero star formation/inflow/outflow rates within
a given halo mass bin. The weights used in the average ($w_i$) are set to the inverse
of the total number of galaxies identified at that snapshot, such that
each snapshot contributes equally to the average.

In addition to the various mass exchanges described previously, we have also
measured additional terms including the mass that is ejected/reaccreted
from galaxies and haloes that does not pass the velocity cuts described
in Section~\ref{measure_sec}, and is therefore not considered
as a genuine outflow, and can be thought of instead as a combination of
noise as particles fluctuate across the somewhat arbitrary boundaries we define for the ISM
and CGM, and low-velocity fountain flows. We also measure the gas accretion rates onto galaxies and haloes
that has been ejected from non-progenitor galaxies and haloes 
\cite[i.e., ``galactic'' or ``halo'' transfer, e.g.,][]{AnglesAlcazar17, Borrow20, Mitchell20b}.
We include both these ``failed-outflow'' and ``transfer'' terms in the N12 model, averaging them
as a function of halo mass and redshift as with the main model terms.
In practice the transfer terms are simply folded into $G_{\mathrm{acc}}^{\mathrm{gal}}$
and $G_{\mathrm{acc}}^{\mathrm{halo}}$
(and are adjusted accordingly if $G_{\mathrm{acc}}^{\mathrm{gal}}$, $G_{\mathrm{acc}}^{\mathrm{halo}}$ are themselves adjusted).
The failed-outflow terms are assumed to scale linearly with $M_{\mathrm{ISM}}$
at the galaxy scale, and with $M_{\mathrm{CGM}}$ at the halo scale,
such that $\dot{M}_{\mathrm{ISM}} = (G_{\mathrm{out,fail}}^{\mathrm{gal}} - G_{\mathrm{ret,fail}}^{\mathrm{gal}}) \, M_{\mathrm{ISM}} \, /t$,
and $\dot{M}_{\mathrm{CGM}} = (G_{\mathrm{out,fail}}^{\mathrm{halo}} - G_{\mathrm{ret,fail}}^{\mathrm{halo}}) \, F_{\mathrm{CGM}}^{\mathrm{pr}} \, M_{\mathrm{CGM}} \, /t$, 
in addition to the mass exchanges described in Eqn.~\ref{ODE_MATRIX}.
Including these extra terms slightly improves the accuracy of our fiducial model, but
they are generally subdominant compared to the main terms.

\begin{figure}
\includegraphics[width=20pc]{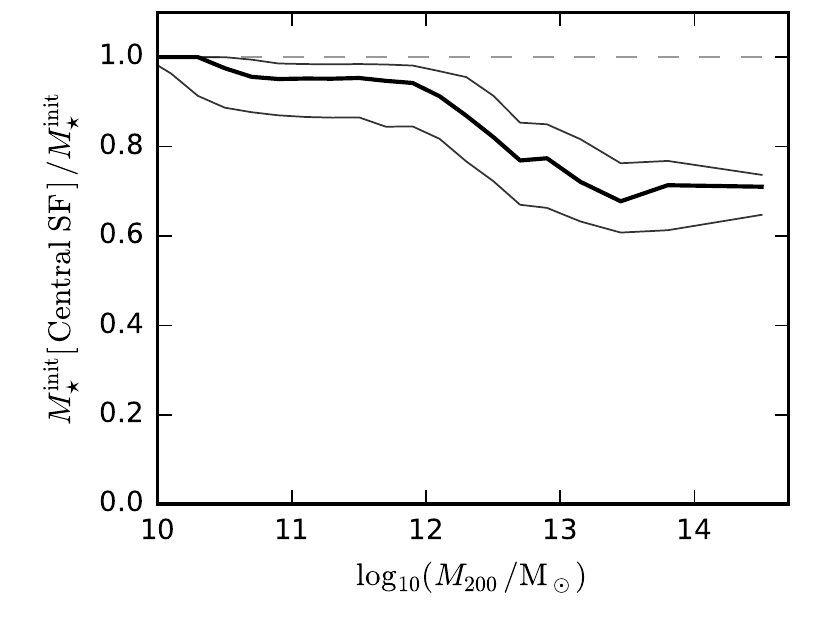}
\caption{The fraction of star formation that occurs within progenitor subhaloes that are
classed as central (rather than as satellites), plotted as a function of halo mass at $z=0$.
Note that this fraction is higher than the ``in situ'' fraction (not plotted), which
includes only stars within the main progenitor subhalo.
Fractions are computed for galaxies that are central at $z=0$, summing over all progenitor
subhaloes.
Solid lines show the $16$, $50$, and $84^{\mathrm{th}}$ percentiles of the distribution.
On average, the total mass formed in central progenitors is between $70 \, \%$ and $100 \, \%$,
with the fraction decreasing with increasing halo mass.
This is relevant to our analysis since any variations to the N12 model will not affect
the stars that are forming within subhaloes that are classified as satellites.
}
\label{cent_sat_frac}
\end{figure}

Unlike \cite{Neistein12}, we have not computed inflow/outflow rates for
satellite galaxies, and as such we compute average values of each term in
Eqn.~\ref{ODE_MATRIX} for central
galaxies only\footnote{Note that we do track the ejection/recycling of particles in \eagle satellite galaxies,
such that they are correctly added to the ejected gas reservoir of a central galaxy
after a merger.}. In our implementation of the N12 model, we evolve
a given satellite galaxy in the model simply by adjusting the mass of each
component to match the change recorded for that galaxy in the underlying
hydrodynamical simulation (this only occurs when the galaxy is classed
as a satellite, the galaxy is evolved as normal as a central before it is accreted onto
the halo of the host). 
A consequence of this strategy is that if we choose to adjust a given term in our
model (away from the fiducial values measured from the underlying simulation), 
the change will only affect the evolution of central galaxies at that snapshot.
We also account for mass exchanges between central
and satellite subhaloes in this manner\footnote{
Such host/satellite mass
  exchanges are subdominant for lower mass subhaloes
  ($M_{200} < 10^{12} \, \mathrm{M_\odot}$), but do appear to make a difference
  for the stellar mass in massive haloes, especially for $M_{200} > 10^{13} \, \mathrm{M_\odot}$.
  We include them in our model to reduce the model bias for massive haloes,
  but we have verified that they do not affect our results, apart from
  in Fig.~\ref{boost_fig}, where they are disabled for that reason.}
, as well as any other possible mass exchanges
that are absent from Eqn.~\ref{ODE_MATRIX} (such as the ejection of
stars from haloes, which can occur in mergers).

To gauge the importance of fixing the evolution of satellites to match
the parent simulation (i.e., not applying Eqn.~\ref{ODE_MATRIX} to
galaxies that are satellites at a given snapshot) for the final stellar mass of galaxies,
Fig.~\ref{cent_sat_frac} shows the fraction of the final stellar mass that
was formed in progenitors while they were classified as centrals (as opposed
to satellites), plotted for descendant galaxies that are centrals at $z=0$.
This fraction decreases from nearly $100 \, \%$ in low-mass central galaxies to $\approx 70 \, \%$
in galaxy groups and clusters. Satellites are negligible for
stellar mass assembly in low-mass haloes, but even in the most massive
haloes (where stellar mass assembly in the central galaxy is dominated by accretion of stars from satellites) most
of the stars were formed in earlier progenitor galaxies while they
were still centrals. As such, the way that we evolve satellites is not crucial, but it should still
be kept in mind that any modifications we make to the model
will have a slightly reduced impact on stellar mass assembly for haloes with mass
$\gg 10^{12} \, \mathrm{M_\odot}$. 

As a final detail, the computation of the quantity $F_{\mathrm{CGM}}^{\mathrm{pr}}$ that appears
in Eqn.~\ref{ODE_MATRIX} requires some additional steps. We introduce this complexity
in order to be able to cleanly separate first-time versus recycled gas accretion
onto galaxies and haloes. Specifically, we want to ensure that increasing
the efficiency of galactic outflows ($\eta^{\mathrm{gal}}$) in our model
does not implicitly increase the rate of first-time gas infall from the CGM.
From the Reference \eagle simulation, we compute two terms in addition to
those shown in Fig.~\ref{coeff_fig}, representing the probability that
halo-scale outflows and gas recycling includes gas that was ejected from the
ISM in the past. Full details are presented in Appendix~\ref{ap_bias_plots}.

\subsection{Measured coefficients}

\begin{figure*}
\begin{center}
\includegraphics[width=40pc]{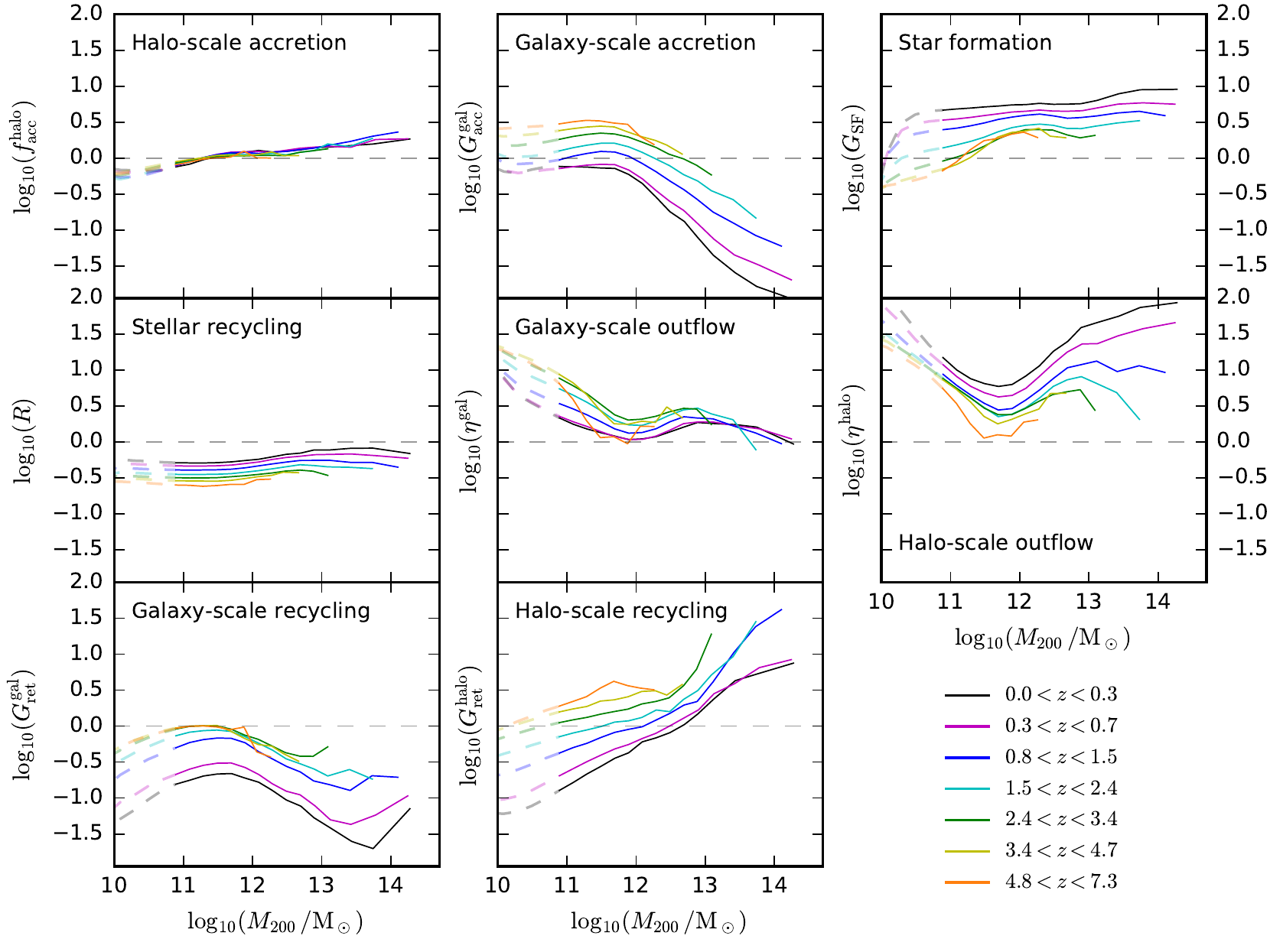}
\caption{An overview of the various dimensionless coefficients that appear in our version of the N12 model \protect (Eqn.~\ref{ODE_MATRIX}),
  plotted as a function of halo mass.
  Each panel corresponds to a given coefficient, and each line colour corresponds to a specific
  redshift bin (as labelled in the bottom-right corner).
  A common x and y-axis range is used for all panels. 
  Solid (dashed) lines indicate the halo mass range for which galaxies in the underlying
  reference \eagle simulation contain on average more (fewer) than one hundred stellar particles. 
}
\label{coeff_fig}
\end{center}
\end{figure*}

Fig.~\ref{coeff_fig} shows the range of averaged coefficients that
appear in our implementation of the N12 model, excluding the highest
redshift bin for visual clarity. As discussed in \cite{Mitchell20b,Mitchell20a},
most of the coefficients shown in Fig.~\ref{coeff_fig} 
exhibit a dependence on halo mass that will preferentially suppress the formation of stars
in either (or both) low-mass haloes ($M_{200} \ll 10^{12} \, \mathrm{M_\odot}$) and
high-mass haloes ($M_{200} \gg 10^{12} \, \mathrm{M_\odot}$). 

Halo-scale first time gas accretion ($f_{\mathrm{acc}}^{\mathrm{halo}}$, 
upper-left panel) increases in efficiency with increasing halo mass,
and will therefore suppress star formation (relatively) in low-mass haloes.
As discussed in \cite{Mitchell20b,Wright20}, feedback reduces first-time
gas accretion rates at $R_{200}$ for $M_{200} < 10^{11} \, \mathrm{M_\odot}$ in \eagle,
but this delayed early accretion is compensated for by increased
first time accretion rates onto more massive haloes (that are naturally
the descendants of the lower-mass haloes), in excess
of the simple expectation following from first time dark matter accretion rates.
Galaxy-scale first time gas accretion ($G_{\mathrm{acc}}^{\mathrm{gal}}$, 
upper-middle panel) drops sharply in efficiency for 
$M_{200} > 10^{12} \, \mathrm{M_\odot}$ (due to inefficient radiative cooling
and the impact of AGN feedback), and will therefore suppress star
formation in the most massive haloes. The efficiency of star formation
per unit ISM mass ($G_{\mathrm{SF}}$, upper-right panel) generally increases 
with increasing halo mass, implying that star formation is preferentially
suppressed in low-mass haloes (we will show later however that $G_{\mathrm{SF}}$
does not shape the SHM relation in the Reference \eagle simulation).

The galaxy- and halo-scale outflow mass loading factors ($\eta^{\mathrm{gal}}$, 
$\eta^{\mathrm{halo}}$, central and middle-right panels) depend negatively on
halo mass for $M_{200} < 10^{12} \, \mathrm{M_\odot}$, reflecting the relative
ease with which stellar feedback can eject baryons from low-mass galaxies and
haloes. For $M_{200} > 10^{12} \, \mathrm{M_\odot}$, the mass loading factors
instead increase with increasing halo mass (especially for $\eta^{\mathrm{halo}}$),
due to AGN feedback. Gas ejection via outflows should therefore suppress star
formation in both low and high-mass haloes. Note that $\eta^{\mathrm{halo}}$
is generally larger than $\eta^{\mathrm{gal}}$ at low redshift 
\cite[see][for a discussion]{Mitchell20a}. On the other hand, $\eta^{\mathrm{gal}}$
more directly affects the ISM, and so it is not obvious which of these two terms
is more important for the SHM relation.

Finally, galaxy-scale wind recycling ($G_{\mathrm{ret}}^{\mathrm{gal}}$, lower-left panel)
peaks in efficiency at $M_{200} \approx 10^{12} \, \mathrm{M_\odot}$, and could
enhance star formation at this mass scale.
Halo-scale wind recycling ($G_{\mathrm{ret}}^{\mathrm{halo}}$, lower-middle panel)
increases monotonically in efficiency with increasing halo mass, and would be expected therefore
to tilt the SHM relation in towards a more positive slope.

It is also apparent that the redshift evolution of the dimensionless terms can vary
in sign from one process to another: for example the dimensionless
infall efficiency ($G_{\mathrm{acc}}^{\mathrm{gal}}$) decreases with decreasing redshift,
whereas the dimensionless star formation efficiency ($G_{\mathrm{SF}}$)
increases with decreasing redshift. It is not obvious how these
opposing trends will combine to produce the resulting redshift evolution
of the SHM relation.

\section{Basic expectations for the SHM relation}
\label{dave_sec}

Before applying the full N12 model, we first use the measurements of inflow and outflow
rates in \eagle to outline a basic expectation for the relative importance
of ``preventative feedback'', galactic outflows, and wind recycling in shaping the 
mass dependence of the SHM relation. Specifically, if we ignore the flows within
the circum-galactic medium, we can reduce the relevant equations to the following.

First, we can assume that first-time gas accretion onto the ISM of galaxies tracks the
accretion of gas at the halo virial radius, such that the galactic first-time accretion
rate is given by 
$\dot{M}_{\mathrm{ISM}} = f_{\mathrm{prev}} \, f_{\mathrm{b}} \dot{M}_{200}$, where
$f_{\mathrm{b}} \equiv \frac{\Omega_{\mathrm{b}}}{\Omega_{\mathrm{m}}}$ is the cosmic baryon fraction.
Here, $f_{\mathrm{prev}}$ represents the combined effects of gravitational heating and feedback
in reducing the accretion rates of gas onto galaxies, referred to collectively as preventative
feedback. 

Second, if we make the assumption that galaxy formation is self-regulated \cite[e.g.,][]{Finlator08,Schaye10,Dave12,Lilly13},
such that the gas accretion rate onto the ISM balances the star formation plus outflow rate (neglecting
stellar mass loss for simplicity),
then the resulting mass conservation equation is

\begin{equation} 
\dot{M}_{\star} + \dot{M}_{\mathrm{out}} = (1 +\eta^{\mathrm{gal}}) \dot{M}_{\mathrm{\star}} = f_{\mathrm{prev}} f_{\mathrm{b}} \dot{M}_{200} + \dot{M}_{\mathrm{ret}},
\label{m14_eq1}
\end{equation}

\noindent where $\dot{M}_{\mathrm{out}}$ is the galaxy-scale outflow
rate, $\eta^{\mathrm{gal}}$ is the galaxy-scale mass loading factor,
and $\dot{M}_{\mathrm{ret}}$ is the rate of recycled gas accretion.
As in Eqn.~\ref{ODE_MATRIX}, we can assume that galactic wind recycling can be parametrised as 
$\dot{M}_{\mathrm{ret}} = G_{\mathrm{ret}}^{\mathrm{gal}} \, M_{\mathrm{ej}}^{\mathrm{gal}} / t$, where $G_{\mathrm{ret}}^{\mathrm{gal}}$
is the characteristic gas return efficiency, $t$ is the age of the Universe at a given redshift,
and $M_{\mathrm{ej}}^{\mathrm{gal}}$ is the mass in the gas reservoir that has been ejected 
from the ISM. This mass is in turn given by the integral 
$M_{\mathrm{ej}}^{\mathrm{gal}} = \int ( \eta^{\mathrm{gal}} \dot{M}_\star - \dot{M}_{\mathrm{ret}} ) \, \mathrm{d}t$.
By substituting in Eqn.~\ref{m14_eq1}, we have
$M_{\mathrm{ej}}^{\mathrm{gal}} = \int \left[ \frac{\eta^{\mathrm{gal}}}{1+\eta^{\mathrm{gal}}}( f_{\mathrm{prev}} f_{\mathrm{b}} \dot{M}_{200} + \dot{M}_{\mathrm{ret}} )  - \dot{M}_{\mathrm{ret}} \right] \, \mathrm{d}t$.
If we make the approximations that $\eta^{\mathrm{gal}} \gg 1$ then this reduces to
$M_{\mathrm{ej}}^{\mathrm{gal}} = \int f_{\mathrm{prev}} f_{\mathrm{b}} \dot{M}_{200} \, \mathrm{d}t$, which
further reduces to $M_{\mathrm{ej}}^{\mathrm{gal}} = f_{\mathrm{prev}} f_{\mathrm{b}} \dot{M}_{200} \, t$ if
$\dot{M}_{200}$ and $f_{\mathrm{prev}}$ are assumed constant as a galaxy evolves. Substituting this back
into Eqn.~\ref{m14_eq1} yields

\begin{equation}
\dot{M}_{\star} \sim \frac{1}{(1+\eta^{\mathrm{gal}})} \left(1 + G_{\mathrm{ret}}^{\mathrm{gal}}\right)  f_{\mathrm{prev}} f_{\mathrm{b}} \dot{M}_{200}.
\end{equation}

\noindent While the approximations used are somewhat crude, the resulting expression has the advantage that the differential effects of
galactic outflows ($\eta^{\mathrm{gal}}$), preventative feedback ($f_{\mathrm{prev}}$), 
and wind recycling ($G_{\mathrm{ret}}^{\mathrm{gal}}$)
are cleanly separated. If we further make the approximation that these three terms are constant as galaxies evolve, then it follows finally that 

\begin{equation}
\frac{M_{\star}}{M_{200}} \sim f_{\mathrm{prev}} f_{\mathrm{b}} \frac{1}{(1+\eta^{\mathrm{gal}})} \left(1 + G_{\mathrm{ret}}^{\mathrm{gal}}\right).
\label{dave_eq}
\end{equation}

\begin{figure}
\includegraphics[width=20pc]{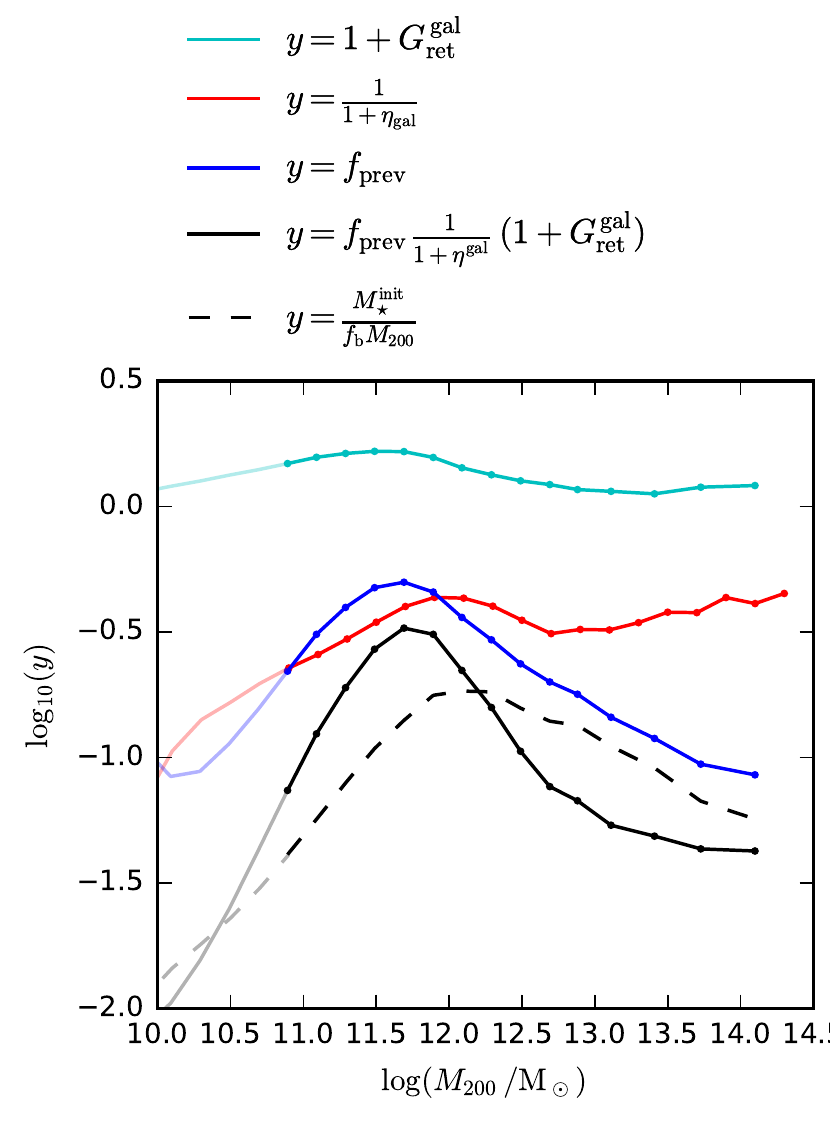}
\caption{
A simplified view of the relative importance of preventative feedback due to inefficient radiative cooling and outflows (blue),
ejection of the ISM by galaxy-scale outflows (red), and galaxy-scale wind recycling (cyan) in shaping
the overall SHM relation (black).
Specifically, subsets of terms from \protect Eqn.~\ref{dave_eq} are plotted as a function of
halo mass, with the solid black line showing the combined expression, which represents
an approximate prediction for the shape of the SHM relation.
The dashed black line shows the actual SHM relation from \eagle for reference.
\protect Eqn.~\ref{dave_eq} ignores stellar recycling, and so we show stellar masses from \eagle that
do not include stellar mass loss ($M_{\star}^{\mathrm{init}}$).
For each line, lighter shades indicate the halo mass range for which galaxies contain on average fewer than $100$
star particles.
All terms and masses are plotted given their measured values averaged over $0.8 < z < 1.5$.
Taken at face value, the preventative feedback term $f_{\mathrm{prev}}$ (blue line) is primarily
responsible for shaping the SHM relation, with ISM ejection (red line) playing a lesser but still
important role, and with wind recycling (cyan) also playing an again lesser (but not negligible)
role. 
We will revisit this basic picture using the full N12 model in \protect Section~\ref{results_sec}.
}
\label{dave_fig}
\end{figure}

Fig.~\ref{dave_fig} shows how the different terms in Eqn.~\ref{dave_eq} add multiplicatively to give
an approximate prediction for the SHM relation. We choose to show the terms (and the \eagle SHM) 
evaluated at $z \approx 1$, which is approximately the mid point for the evolution that each term undergoes
at fixed halo mass 
(see Fig.~\ref{coeff_fig}). In addition, since Eqn.~\ref{dave_fig} ignores stellar recycling, we
show the \eagle SHM relation for stellar masses before any stellar mass loss (i.e., we sum
the initial mass of each star particle in a galaxy).

Considering first the term that relates to wind recycling 
($1+G_{\mathrm{ret}}^{\mathrm{gal}} \equiv 1+t/\tau_{\mathrm{ret}}^{\mathrm{gal}}$, cyan line),
this term saturates when $G_{\mathrm{ret}}^{\mathrm{gal}} \ll 1$ ($\tau_{\mathrm{ret}}^{\mathrm{gal}} \gg t$), 
since wind recycling is of little importance
if the material ejected in galactic winds takes on average longer than the age of the Universe to return.
As such, while $G_{\mathrm{ret}}^{\mathrm{gal}}$ itself clearly follows the approximate shape of the SHM relation
(see Fig.~\ref{coeff_fig}, lower-left panel), the long recycling timescales in \eagle are expected
to blunt the actual impact of wind recycling on the shape of the SHM relation. Recycling is more
efficient (relative to the Hubble time) at high redshift, and so may play a larger role in this regime.
Even for the values at $z\approx 1$ shown in Fig.~\ref{dave_fig}, there is however still an appreciable
curvature in the cyan line that contributes to the peak of the SHM ratio at a halo mass slightly
below $10^{12} \, \mathrm{M_\odot}$.

Considering second the preventative feedback term ($f_{\mathrm{prev}}$, blue line in Fig.~\ref{dave_fig}),
it is clearly implied that this term is the most important (over ISM ejection via outflows and wind recycling) 
for setting the overall shape of the SHM relation. $f_{\mathrm{prev}}$ peaks strongly at a halo
mass slightly below $10^{12} \, \mathrm{M_\odot}$, and its effect does not 
saturate for low values of $f_{\mathrm{prev}}$, since first-time inflows represent
the source term for galactic gaseous assembly.

Thirdly, the ISM ejection outflow term ($1/(1+\eta^{\mathrm{gal}})$, red line in Fig.~\ref{dave_fig}) has
an apparent intermediate level of importance between the wind recycling and preventative feedback 
terms. This term peaks at $M_{200} \sim 10^{12} \, \mathrm{M_\odot}$, at a slightly higher
halo mass relative to the other two terms. 

Combining these terms together (solid black line in Fig.~\ref{dave_fig}), Eqn.~\ref{dave_eq}
predicts an SHM that is qualitatively similar to that of \eagle (dashed black line), but
with various quantitative differences. Relative to \eagle, Eqn.~\ref{dave_eq} predicts an SHM that peaks
at slightly lower halo masses, that has a steeper slope for halo masses below the peak mass, and that is
more sharply peaked. These differences stem from the fact that we are ignoring the
redshift evolution that each term undergoes, from the simplifying approximations invoked
to derive Eqn.~\ref{dave_eq}, and from the fact that we are ignoring accretion of satellite galaxies
and subhaloes (which affect the galaxy stellar mass and halo mass respectively). Ignoring satellite galaxies
and subhaloes is expected to be a poor approximation for high-mass haloes, where accreted stars 
formed ex-situ are thought to form a large fraction
of the total stellar mass \cite[e.g.,][]{deLucia07,Moster18}, which is indeed the case
for \eagle \cite[][]{Qu17,Clauwens18}.

Despite these simplifications, Eqn.~\ref{dave_eq} does clearly capture the basic behaviour. 
That said, the ``preventative feedback'' term in Eqn.~\ref{dave_eq} effectively
combines most of the gas flows that affect the circum-galactic medium into a single
term. In the framework of the equilibrium model of \cite{Dave12}, $f_{\mathrm{prev}}$
conceptually represents the effects of feedback processes in slowing the first-time
accretion rates of gas onto galaxies. As we shall show in the following section,
once $f_{\mathrm{prev}}$ is split into different (time and mass-dependent) parts, including a reduction of gas inflow rates
at the virial radius ($f_{\mathrm{acc}}^{\mathrm{halo}}$ in Eqn.~\ref{ODE_MATRIX}), first-time
infall from the CGM to the ISM ($G_{\mathrm{acc}}^{\mathrm{gal}}$), halo-scale outflows ($\eta^{\mathrm{halo}}$),
and halo-scale wind recycling ($G_{\mathrm{ret}}^{\mathrm{halo}}$), it becomes apparent that
(of these terms) it is the ejection of gas beyond the halo virial radius combined with inefficient
gas infall in very massive haloes that is primarily responsible for shaping the SHM relation
in \eagle.

\section{Results}
\label{results_sec}

\begin{figure}
\includegraphics[width=20pc]{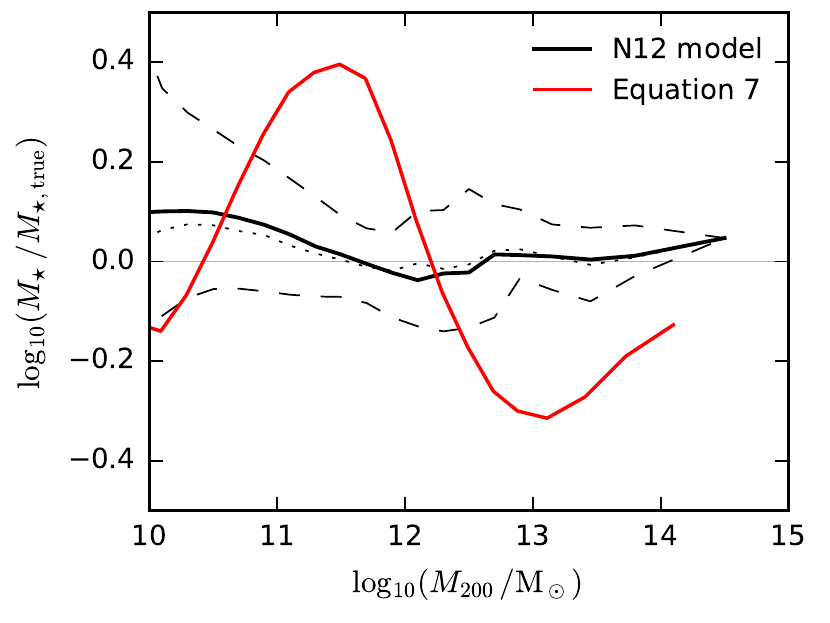}
\caption{
  The residual in stellar mass between the N12 model and \eagle at $z=0$,
  plotted for central galaxies as a function of halo mass.
  Solid black lines show the median, dashed lines show the $16$ and $84^{\mathrm{th}}$ percentiles of the distribution,
  and the dotted line shows the mean residual.
  Overall, the N12 model reproduces \eagle stellar masses to within a few tens of
  percent; the largest deviation is seen in very low-mass haloes, where the N12 model
  over-predicts the stellar mass by $\approx 30 \, \%$.
  For comparison, the residual from the simplified model discussed in Section~\ref{dave_sec},
    given by Eqn.~\ref{dave_eq}, is shown as a red solid line.
  As in Fig.~\ref{dave_fig} (solid black line), Eqn.~\ref{dave_eq} is evaluated for
  model terms measured at $z \approx 1$.
}
\label{accuracy_fig}
\end{figure}

Fig.~\ref{accuracy_fig} demonstrates the accuracy with which the fiducial implementation of
the N12 model is able
to reproduce the SHM relation from the original reference \eagle simulation. The
two agree to within $10 \, \%$ on average for $M_{200} > 10^{12} \, \mathrm{M_\odot}$. Agreement
is worse at lower halo masses, with the model biased slightly high by $\approx 30 \, \%$
for $M_{200} < 10^{11} \, \mathrm{M_\odot}$.

\subsection{What sets the normalisation of the SHM relation?}
\label{norm_subsec}

\begin{figure*}
\begin{center}
\includegraphics[width=40pc]{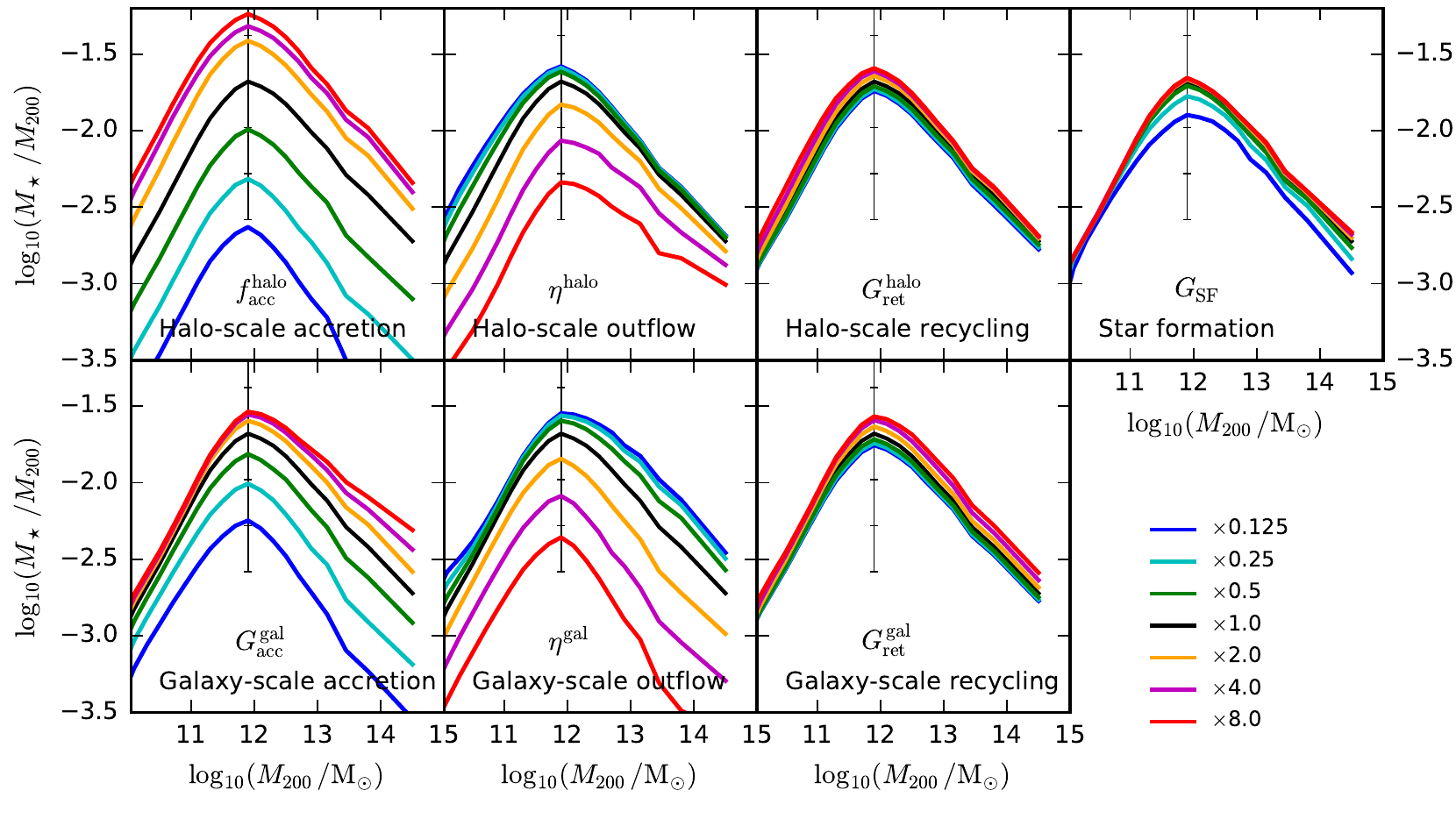}
\caption{The impact of multiplying individual terms in our implementation of the N12 model by a constant 
\protect (see Eqn.~\ref{ODE_MATRIX}). 
The ratio of galaxy stellar mass to halo mass is plotted as a function of halo mass
for central galaxies at $z=0$.
The black line shows the fiducial model, and coloured lines show the
model with a single term multiplied by the factor indicated in the legend.
Each panel corresponds to a different term being modified.
Note that for the first-time halo-scale accretion term ($f_{\mathrm{acc}}^{\mathrm{halo}}$, upper-left panel),
we cap modifications such that haloes cannot accrete more than the maximum recorded mass-averaged rate in
\eagle at a given redshift, to avoid situations where halo baryon content exceeds the universal baryon fraction.
The horizontal notches connected to vertical grey lines indicate how the SHM 
relation would change if there were a linear response to the multiplying constant.
There are clear differences in the relative importance of the different terms in the model. 
For example, changing the outflow efficiency terms
has a much larger effect than changing the star formation and recycling terms.
}
\label{boost_fig}
\end{center}
\end{figure*}

As a first application of the N12 model, we can simply multiply a single
term that appears in Eqn.~\ref{ODE_MATRIX} by a constant factor, and thereby gauge
the relevant importance of that term to the normalisation of the SHM relation. We consider both positive
and negative factors of two, four and eight\footnote{
Since the changes considered can be quite extreme, we choose here to disable the 
mass evolution of satellite galaxies (i.e., their star formation, inflow, outflow
rates are all set to zero), which otherwise are evolved to exactly 
follow changes of the original satellites from \eagle (see Section~\ref{n12_desc_sec}). 
Without this change, star formation within satellites can in some cases become the 
dominant mode of galaxy stellar mass growth (even for central galaxies due to mergers), 
obfuscating our interpretation of the effect of changing a model term.
}. 
The resulting SHM relations for central galaxies are presented in Fig.~\ref{boost_fig}.
For reference, grey ticks show the expected change in the SHM relation if changes in galaxy
stellar mass would scale linearly with changes to the model terms.
This is approximately the case for the first-time gas accretion term ($f_{\mathrm{acc}}^{\mathrm{halo}}$, upper-left panel)
since this acts as the source term (via the term 
$f_{\mathrm{acc}} \, \frac{\Omega_{\mathrm{b}}}{\Omega_{\mathrm{m}}-\Omega_{\mathrm{b}}} \, \dot{M}_{\mathrm{DM}}^{\mathrm{1st}}$
in the expression for $\dot{M}_{\mathrm{CGM}}$, see Eqn.~\ref{ODE_MATRIX}). 
For positive modifications to $f_{\mathrm{acc}}^{\mathrm{halo}}$,
we choose to saturate $f_{\mathrm{acc}}^{\mathrm{halo}}$ at the maximum recorded value for a given redshift bin,
in order to prevent a scenario where haloes contain significantly more than than their share of the cosmic
baryon fraction.

Modifying any other term in the model produces a clearly sub-linear response in the SHM
relation. In addition, it is evidently easier to reduce the efficiency of the
conversion of baryons into stars than it is to increase it; positive responses
of the SHM relation to a given fractional change in the value of a term in Eqn.~\ref{ODE_MATRIX}
are generally smaller (fractionally) in magnitude
than the response to the corresponding fractional change that produces a negative response.
At $M_{200} \sim 10^{12} \, \mathrm{M_\odot}$, this can be partly explained as a saturation
in the maximum possible conversion of baryons into stars; increasing the SHM normalization
by a factor ten would result in over-shooting the cosmic baryon fraction ($f_{\mathrm{b}} = 0.16$)
at this mass scale.
This is clearly not the entire explanation however, since (for example) no model variation 
produces a value of $\log_{10}(M_\star / M_{200}) > -3$ for 
$M_{200} \sim 10^{10} \, \mathrm{M_\odot}$, which is well short of the cosmic baryon fraction.

\subsubsection{Outflows}

We consider first the two outflow terms 
($\eta^{\mathrm{gal}}$, $\eta^{\mathrm{halo}}$, second column panels), which are
the dimensionless mass loading factors defined at the galaxy scale (gas leaving the ISM) 
and at the halo scale (gas moving beyond $R_{200}$). Increasing $\eta^{\mathrm{gal}}$
or $\eta^{\mathrm{halo}}$ produces the strongest response in the SHM relation, when compared
to other model terms. The response is still sub-linear however:
increasing $\eta^{\mathrm{gal}}$ by a factor two reduces galaxy stellar masses
by $\approx 30 \, \%$, a factor of $1.5$. 
For the galaxy-scale outflow, the basic expectation is that $\dot{M}_{\star} \propto \frac{1}{1+\eta^{\mathrm{gal}}}$ 
(i.e. gas in the ISM either goes into stars or into an outflow).
This only elicits a linear response if $\eta^{\mathrm{gal}} \gg 1$, which is not
the case over the entire evolution of a given galaxy (see Fig.~\ref{coeff_fig}).
In addition, while increasing $\eta^{\mathrm{gal}}$ in isolation does reduce the 
star formation rate, the reduced star formation in turn decreases the outflow rate 
at the virial radius, which in turn will result in enhanced rates of first-time gas 
accretion onto the ISM. 

This can be further generalised to consider gas accretion onto the ISM.
In the regime where all of the gas in the CGM will be accreted onto the ISM
within a Hubble time, or will be ejected, we can make the approximation
that $\dot{M}_{\mathrm{acc}}^{\mathrm{halo}} \approx \dot{M}_{\mathrm{acc}}^{\mathrm{gal}} + \dot{M}_{\mathrm{CGM,out}}$,
where $\dot{M}_{\mathrm{acc}}^{\mathrm{halo}}$ is the gas accretion rate onto the halo,
$\dot{M}_{\mathrm{acc}}^{\mathrm{gal}}$ is the gas accretion rate from the CGM onto
the ISM, and $\dot{M}_{\mathrm{CGM,out}}$ is the CGM outflow rate.
If we apply a similar approximation to the ISM 
($\dot{M}_{\mathrm{acc}}^{\mathrm{gal}} \approx (1+\eta^{\mathrm{gal}}) \dot{M}_\star$),
it follows that 
$\dot{M}_{\mathrm{acc}}^{\mathrm{halo}} \approx \dot{M}_\star (1+\eta^{\mathrm{gal}} + \eta^{\mathrm{halo}})$,
since halo-scale outflows are assumed to be proportional to the star formation rate
in our model. If these approximations hold, we see that
changing $\eta^{\mathrm{halo}}$ will only yield a linear response in $\dot{M}_\star$
if $\eta^{\mathrm{halo}} \gg 1 + \eta^{\mathrm{gal}}$ (and similarly for $\eta^{\mathrm{gal}}$).
In other words, it does not matter for the SHM relation what value $\eta^{\mathrm{halo}}$ takes
if the vast majority of the gas accreted onto galaxies is ejected in 
a galaxy-scale outflow, such that the star formation rate (and therefore
the halo outflow rate) is very
small compared to the accretion rate onto the ISM.
In \eagle, $\eta^{\mathrm{halo}} \gg 1 + \eta^{\mathrm{gal}}$ is generally
satisfied at low redshift, but not for for $z>1$ (Fig.~\ref{coeff_fig}). 

Furthermore, reducing the values of either outflow term will essentially produce no response 
if the mass loading factors are $\ll 1$, because in that 
limit gas consumption rather than outflows limits the fuel for star formation.
This is indeed seen in Fig.~\ref{boost_fig}, where the SHM relation saturates
to a constant at a given halo mass as $\eta^{\mathrm{halo}}$ and $\eta^{\mathrm{gal}}$
are reduced significantly below their fiducial values.

\subsubsection{Star formation}

For the model terms that are related to timescales 
($G_{\mathrm{acc}}^{\mathrm{gal}}$, $G_{\mathrm{SF}}$, 
$G_{\mathrm{ret}}^{\mathrm{gal}}$, $G_{\mathrm{ret}}^{\mathrm{halo}}$), their
relative importance to galaxy stellar mass growth will generally depend on whether they represent
a significant bottleneck relative to the Hubble time.
For example, recall that $G_{\mathrm{SF}} \equiv t \, \dot{M}_\star / M_{\mathrm{ISM}} \equiv t / \tau_{\mathrm{SF}}$,
where $t$ is the age of the Universe and $\tau_{\mathrm{SF}}$ is the conventional
ISM depletion timescale. The exact value of $G_{\mathrm{SF}}$
is unimportant for galaxy star formation rates if $G_{\mathrm{SF}} \gg 1$ (i.e. if the ISM
depletion time is much shorter than the Hubble time), which is generally the
case for high-mass galaxies (Fig.~\ref{coeff_fig}). 
Furthermore, the effective ISM depletion timescale
is shortened by $1/(1+\eta^{\mathrm{gal}})$ due to galactic outflows, especially for
low-mass galaxies where $\eta^{\mathrm{gal}} \gg 1$. As such, Fig.~\ref{boost_fig} shows
that the only case for
which the value of $G_{\mathrm{SF}}$ has any impact on the SHM relation is when 
$G_{\mathrm{SF}}$ is reduced to $\approx 10 \, \%$ of its fiducial value, at which
point a given hydrogen atom accreted onto the halo would on average have
to spend a significant fraction of a Hubble time in the ISM before being locked
into a forming star. 

\subsubsection{Galaxy-scale accretion}

On similar grounds, we expect that the exact value of the galaxy-scale accretion efficiency term
$G_{\mathrm{acc}}^{\mathrm{gal}}$ would be unimportant if it is $\gg 1$. As a reminder, this 
term represents the efficiency with which circum-galactic gas that has not yet 
been part of the galaxy's ISM is able to accrete onto the ISM, defined relative to 
the Hubble time. In low-mass haloes ($M_{200} < 10^{12} \, \mathrm{M_\odot}$) where the virial temperature is not high
enough to push beyond the peak of the radiative cooling curve,
the basic expectation is that gravitational timescales of order the halo
dynamical time limit gas infall
from the CGM to the ISM. Since the dynamical time scales as $1/\sqrt{\rho}$
and halo densities exceed the cosmic mean by $\sim 10^2$, this would
give $G_{\mathrm{infall}} \sim 10$. 
As discussed in \cite{Mitchell20b}, feedback processes in \eagle inhibit gas
infall, pushing $G_{\mathrm{acc}}^{\mathrm{gal}}$ to lower values \cite[Fig.~\ref{coeff_fig}; see also, e.g.,][]{Nelson15,Pandya20}, 
which can even be less than unity for low-mass galaxies at $z<1$ (i.e. the average infall time is longer
than the Hubble time at low redshift). For high-mass haloes, long characteristic
radiative timescales plus AGN feedback push $G_{\mathrm{acc}}^{\mathrm{gal}}$ to very low values
(one of the requirements for creating passive central galaxies).

As such, we expect that adjusting $G_{\mathrm{acc}}^{\mathrm{gal}}$ in Fig.~\ref{boost_fig}
by a given factor will yield a larger response than adjusting $G_{\mathrm{SF}}$, which is
indeed the case. The largest response is seen for the most massive haloes,
where the fiducial values of $G_{\mathrm{acc}}^{\mathrm{gal}}$ are lowest. For 
$M_{200} \sim 10^{12} \, \mathrm{M_\odot}$, the effect of increasing $G_{\mathrm{acc}}^{\mathrm{gal}}$ 
quickly saturates as typical values of $G_{\mathrm{acc}}^{\mathrm{gal}}$ are generally
much larger than unity. Decreasing $G_{\mathrm{acc}}^{\mathrm{gal}}$ produces a much
larger fractional response, as expected.

\subsubsection{Wind recycling}

Fig.~\ref{boost_fig} shows that modifying either of the gas
recycling efficiencies ($G_{\mathrm{ret}}^{\mathrm{gal}}$, $G_{\mathrm{ret}}^{\mathrm{halo}}$) elicits
only a weak response in the SHM relation. For recycling timescales
that are very long compared to the Hubble time ($G_{\mathrm{ret}} \ll 1$),
recycling becomes negligible for stellar mass growth, and the exact values
of the recycling terms will be unimportant (see Eqn.~\ref{dave_eq}); the
same is true if the outflow ejection terms ($\eta$) are small.
Fig.~\ref{coeff_fig} shows that the galaxy-scale recycling term ($G_{\mathrm{ret}}^{\mathrm{gal}}$)
is much less than unity at low redshift, but is marginally efficient ($\approx 1$) for
$z \geq 1$, and indeed in \cite{Mitchell20b} we show explicitly that galaxy-scale recycling
in \eagle plays a secondary but still important role for stellar mass assembly.

Decreasing the efficiency of gas recycling in our model is therefore expected to quickly
saturate as recycling becomes negligible. Conversely, increasing $G_{\mathrm{ret}}^{\mathrm{gal}}$
will saturate if this results in $G_{\mathrm{ret}}^{\mathrm{gal}} \gg G_{\mathrm{SF}}$,
as in this case the star formation efficiency $G_{\mathrm{SF}}$ becomes the bottleneck, with recycled gas
spending most of its time in the ISM rather than in the CGM. From Fig.~\ref{coeff_fig},
this will start to be the case for $z > 1.5$, but not at lower redshifts.
In addition, while increasing $G_{\mathrm{ret}}^{\mathrm{gal}}$
does by definition allow more of the gas ejected from the ISM
to participate in later star formation, this also depletes the CGM of galactic wind
material. This in turn means that more of the CGM that was not ejected in a
galactic wind will be ejected outside of the virial radius (i.e. preventative
feedback for first-time infalling gas becomes more effective).

Fig.~\ref{coeff_fig} shows that halo-scale recycling 
($G_{\mathrm{ret}}^{\mathrm{halo}}$) becomes very inefficient in low-mass haloes 
($M_{200} < 10^{12} \, \mathrm{M_\odot}$) at low redshift.
For high-mass haloes, gas accretion onto the ISM ($G_{\mathrm{acc}}^{\mathrm{gal}}$, $G_{\mathrm{ret}}^{\mathrm{gal}}$) 
instead becomes very inefficient, rendering the high efficiency of halo-scale
recycling in this regime somewhat unimportant for stellar mass growth.
In addition, increasing $G_{\mathrm{ret}}^{\mathrm{halo}}$ to arbitrarily large
values will quickly saturate in effect since the galaxy-scale gas accretion timescales 
($G_{\mathrm{acc}}^{\mathrm{gal}}$, $G_{\mathrm{ret}}^{\mathrm{gal}}$) are
always at least comparable to the Hubble time for all halo masses, at which point the time
spent outside $R_{200}$ for a given ejected hydrogen atom is not an important
bottleneck for stellar mass growth, compared to the time spent in the CGM afterwards. 
We have checked that if we artificially boost the efficiencies of gas accretion onto the ISM 
($G_{\mathrm{acc}}^{\mathrm{gal}}$, $G_{\mathrm{ret}}^{\mathrm{gal}}$)
to very high values (such that the time spent outside $R_{200}$ is always the bottleneck),
the SHM relation does respond strongly to the chosen value of $G_{\mathrm{ret}}^{\mathrm{halo}}$.
Under physically sensible conditions, we therefore find that the exact efficiency of halo-scale recycling is not important
for galaxy stellar mass growth (at least within the scenario for gas flows around 
galaxies presented by the \eagle simulation).

\subsubsection{Summary}

Putting this all together, what is important (and what is not important)
for setting the overall normalisation of the SHM relation, for a given
amount of first-time cosmological gas accretion at the virial radius?
In terms of which model terms produce the strongest response,
the mass ejected from the ISM per unit star formation is
important for all galaxies, as is the energy ejected from the ISM
that goes into powering outflows at larger spatial scales (especially
for $M_{200} < 10^{12} \, \mathrm{M_\odot}$).
The efficiency with which gas is accreted for the first time
from the CGM onto the ISM is almost as important (especially for
$M_{200} > 10^{12} \, \mathrm{M_\odot}$). 
The efficiency of gas recycling is less important in comparison (but not negligible),
and finally the chosen value of the star formation efficiency
is inconsequential, unless it drops to very low values that
correspond to gas consumption timescales similar to or greater 
than the age of the Universe.

On the other hand, we can also ask what is important for suppressing
the maximum allowed value of $M_\star \, /M_{200}$ 
(set by $f_{\mathrm{acc}}^{\mathrm{halo}}$) down to the value seen
for the fiducial model at $M_{200} \approx 10^{12} \, \mathrm{M_\odot}$.
Framed in this way, the two wind recycling terms are actually
of comparable importance to the outflow terms and the
galaxy-scale gas accretion term. This is because
none of the terms (except for $f_{\mathrm{acc}}^{\mathrm{halo}}$)
are capable of producing a strong positive response in terms
of $M_\star \, / M_{200}$. For outflows, this is mostly because
varying one term (i.e., galaxy or halo-scale) in isolation
is compensated for by the other term. We will show
in the next subsection that varying both outflow terms at the
same time can elicit a much larger positive response at lower
halo masses ($M_{200} \ll 10^{12} \, \mathrm{M_\odot}$).

\subsection{What sets the shape of the SHM relation?}
\label{shm_shape_subsec}

\begin{figure*}
\begin{center}
\includegraphics[width=40pc]{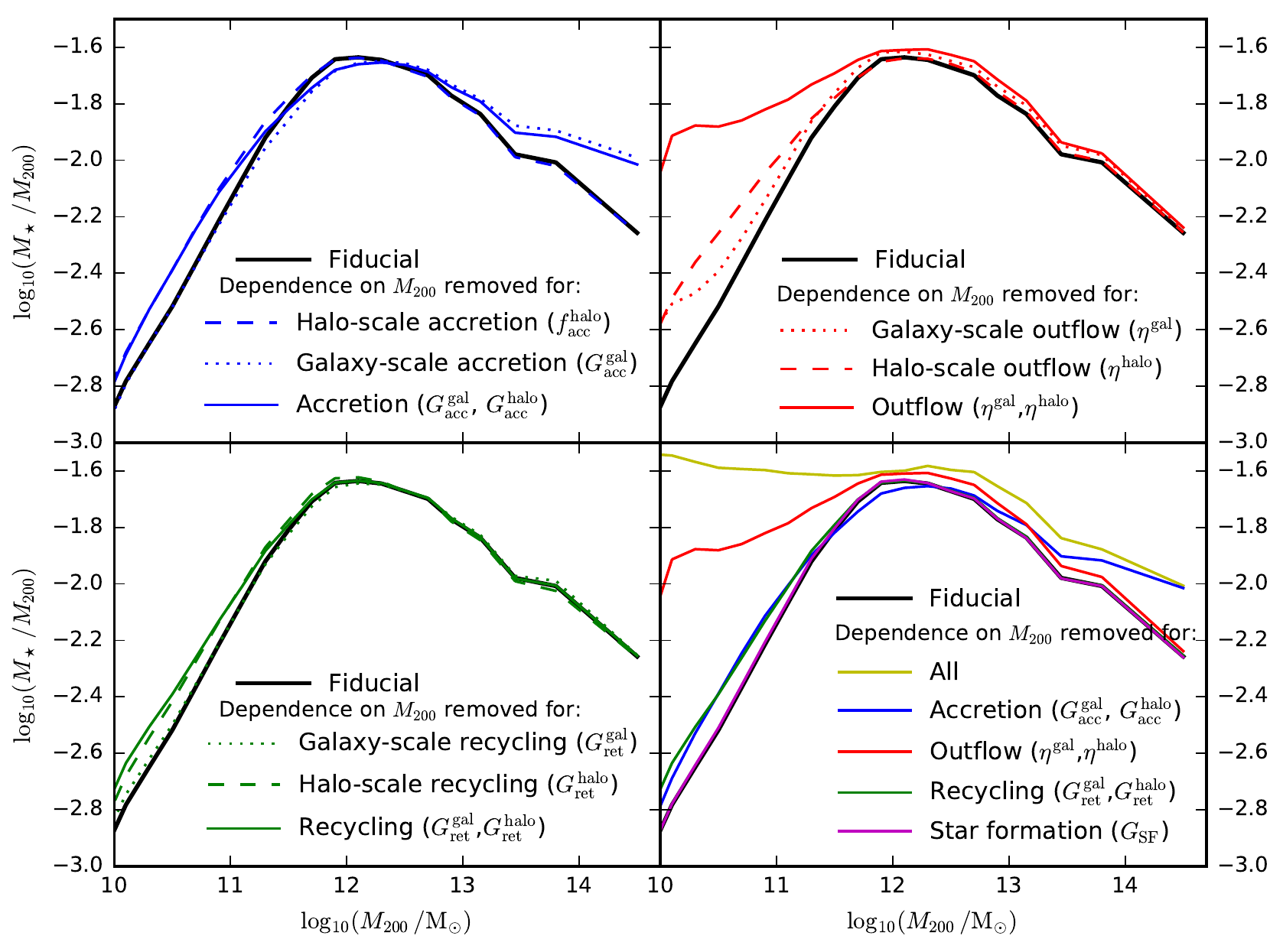}
\caption{The impact of removing the halo mass dependence of one (or more) 
term(s) in the N12 model (see Eqn.~\ref{ODE_MATRIX}), 
by fixing its value(s) to the one(s) corresponding to $M_{200} = 10^{12} \, \mathrm{M_\odot}$
for each redshift bin. 
The median ratio of galaxy stellar mass to halo mass is plotted as a function of halo mass
for central galaxies at $z=0$.
Black lines show the fiducial model and coloured lines show the
model with the term indicated in each panel held fixed.
The ``All'' case (yellow line, bottom-right panel) indicates that all of the considered
modifications are applied (i.e. $f_{\mathrm{acc}}^{\mathrm{halo}}$, $G_{\mathrm{acc}}^{\mathrm{gal}}$, $G_{\mathrm{SF}}$, $\eta^{\mathrm{gal}}$,
$\eta^{\mathrm{halo}}$, $G_{\mathrm{ret}}^{\mathrm{gal}}$, $G_{\mathrm{ret}}^{\mathrm{halo}}$ are all fixed at their respective
values at $M_{200} = 10^{12} \, \mathrm{M_\odot}$).
These modifications indicate how different processes in \eagle shape the dependence of galaxy stellar mass
on halo mass. Roughly speaking, the mass dependence of outflows plays the largest role for 
$M_{200} < 10^{12} \, \mathrm{M_{\odot}}$, but
halo-scale gas accretion (i.e. preventative feedback at $R_{200}$) and halo-scale gas recycling also contribute.
The mass dependence of galaxy-scale gas accretion (i.e., the efficiency of first-time CGM infall onto the ISM) plays
the largest role in shaping the SHM relation for $M_{200} > 10^{12} \, \mathrm{M_{\odot}}$, but outflows
also contribute.
}
\label{m12_fix_fig}
\end{center}
\end{figure*}

Fig.~\ref{m12_fix_fig} shows the impact of removing the halo mass dependence of
terms in the N12 model by fixing them to their values corresponding to the
halo mass $M_{200} = 10^{12} \, \mathrm{M_\odot}$. In doing so,
we can assess the respective role of different terms in setting the shape of
the SHM relation. 

\subsubsection{First-time gas accretion}

The upper-left panel of Fig.~\ref{m12_fix_fig} focuses on first-time inflowing gas, which
is set in the N12 model by $f_{\mathrm{acc}}^{\mathrm{halo}}$ 
(first-time gas accretion at the virial radius) and $G_{\mathrm{acc}}^{\mathrm{gal}}$
(first-time infall from the CGM to the ISM). From Fig.~\ref{coeff_fig},
we can see that $f_{\mathrm{acc}}^{\mathrm{halo}}$ exhibits a shallow and positive dependence
on halo mass, since feedback processes moderately inhibit first-time gas
accretion onto low-mass haloes in \eagle, and this slightly increases the first-time infall
of gas onto more massive descendant haloes as a result \cite[][]{Wright20,Mitchell20b}.
As such, fixing $f_{\mathrm{acc}}^{\mathrm{halo}}$ to its value at $M_{200} = 10^{12} \, \mathrm{M_\odot}$
(dashed blue line in Fig.~\ref{m12_fix_fig}) does not have a dramatic effect on the SHM relation, but does slightly increase
the stellar mass in haloes with $M_{200} < 10^{12} \, \mathrm{M_\odot}$. Interestingly,
fixing $f_{\mathrm{acc}}^{\mathrm{halo}}$ has negligible impact on more massive haloes. This
can be explained by the low galaxy accretion efficiencies ($G_{\mathrm{acc}}^{\mathrm{gal}}$, $G_{\mathrm{ret}}^{\mathrm{gal}}$)
measured for these haloes, such that star formation that occurred in low-mass 
progenitor galaxies (that later merge) dominates the final stellar mass in the central galaxies
of very massive haloes\footnote{
In addition, star formation that occurs in satellites is unaffected by our modifications
to terms in the N12 model, which is a $\approx 30 \, \%$ effect in massive haloes, see Fig.~\ref{cent_sat_frac}.}.
In other words, most of the stars in dark matter haloes with
$M_{200} \gg 10^{12} \, \mathrm{M_\odot}$ are formed from gas that was accreted
for the first time onto less massive progenitor haloes with 
$M_{200} \sim 10^{12} \, \mathrm{M_\odot}$ at earlier times.

As a mirror image to fixing $f_{\mathrm{acc}}^{\mathrm{halo}}$, fixing $G_{\mathrm{acc}}^{\mathrm{gal}}$ 
(dotted blue line in the top-left panel of Fig.~\ref{m12_fix_fig})
does not impact low-mass haloes, but has the strongest impact of any of the model terms
on the stellar mass in haloes with $M_{200} > 10^{12} \, \mathrm{M_\odot}$.
Fig.~\ref{coeff_fig} shows that for $M_{200} < 10^{12} \, \mathrm{M_\odot}$, $G_{\mathrm{acc}}^{\mathrm{gal}}$ exhibits
barely any dependence on halo mass \cite[which can be explained if gas infall
in low-mass haloes is, as expected, connected to gravitational timescales, 
which are scale-free and therefore independent of halo mass; ][]{Mitchell20b}, and as such
does not affect the shape of the SHM relation in this regime.
For $M_{200} > 10^{12} \, \mathrm{M_\odot}$, $G_{\mathrm{acc}}^{\mathrm{gal}}$ decreases
sharply with increasing halo mass as longer radiative cooling times and AGN
feedback take effect, and as such fixing $G_{\mathrm{acc}}^{\mathrm{gal}}$ 
to its value for $M_{200} = 10^{12} \, \mathrm{M_\odot}$ significantly
increases galaxy stellar masses in high-mass haloes. Fixing both $G_{\mathrm{acc}}^{\mathrm{gal}}$
and $f_{\mathrm{acc}}^{\mathrm{halo}}$ simultaneously (solid blue line) has little additional impact, since
the halo mass dependence of the two terms affect the SHM relation in different
regimes.

\subsubsection{Outflows}

Removing the halo mass dependence of the mass loading factors for galaxy or halo-scale
outflows ($\eta^{\mathrm{gal}}$ and $\eta^{\mathrm{halo}}$, respectively the 
dotted and dashed red lines in the upper-right panel of Fig.~\ref{m12_fix_fig}) impacts the 
shape of the SHM relation for $M_{200} < 10^{12} \, \mathrm{M_\odot}$, which
is expected since both are negative functions of halo mass in this range (see Fig.~\ref{coeff_fig}).
The halo mass dependence of the mass loading factors is steeper for halo-scale outflows 
($\eta^{\mathrm{halo}}$), and so accordingly fixing this to its value
at $M_{200} = 10^{12} \, \mathrm{M_\odot}$ has a larger effect.
Interestingly, fixing both $\eta^{\mathrm{gal}}$ and $\eta^{\mathrm{halo}}$
simultaneously (solid red line) has a much larger effect than fixing
one of the two, for $M_{200} < 10^{12} \, \mathrm{M_\odot}$. As discussed in Section~\ref{norm_subsec}, 
decreasing one of $\eta^{\mathrm{gal}}$ or $\eta^{\mathrm{halo}}$ in isolation
actually has the effect of increasing the importance of the other
term (for example, decreasing $\eta^{\mathrm{gal}}$ in isolation increases $\dot{M}_\star$,
which in turn increases the halo-scale outflow rate, which in turn
will decrease the rate of gas infall onto the ISM).
For $M_{200} > 10^{12} \, \mathrm{M_\odot}$, fixing either or
both of $\eta^{\mathrm{gal}}$ and $\eta^{\mathrm{halo}}$ has little
effect. $\eta^{\mathrm{gal}}$ depends only weakly on halo mass
in this range; $\eta^{\mathrm{halo}}$ has a stronger mass dependence
but, as discussed earlier, in this halo mass range gas infall
from the CGM onto the ISM is extremely inefficient (and in any case halo-scale
recycling is very efficient in this regime), rendering halo-scale outflows
irrelevant for stellar mass growth.

\subsubsection{Recycling}

Removing the halo mass dependence of the efficiencies of galaxy or halo-scale gas recycling
($G_{\mathrm{ret}}^{\mathrm{gal}}$ and $G_{\mathrm{ret}}^{\mathrm{halo}}$, respectively
the dotted and dashed green lines in the bottom-left panel of Fig.~\ref{m12_fix_fig})
impacts the SHM relation for $M_{200} < 10^{12} \, \mathrm{M_\odot}$,
but not at higher halo masses. In the low-halo mass
regime, both the galaxy- and halo-scale recycling efficiencies
increase with halo mass (see Fig.~\ref{coeff_fig}), such that fixing them
to their values at $M_{200} = 10^{12} \, \mathrm{M_\odot}$
increases galaxy stellar masses, softening the slope of
the SHM relation.

\subsubsection{Summary \& discussion}

Putting all these components together, the lower-right panel of
Fig.~\ref{m12_fix_fig} shows respectively how the mass dependencies
of outflows, (first-time) accretion, and recycling shape the SHM relation in \eagle.
The yellow line shows the effect of fixing all of these terms
simultaneously to their values at $M_{200} = 10^{12} \, \mathrm{M_\odot}$, 
such that there is effectively no explicit halo-mass dependence in the N12 model.
Accordingly, the SHM relation becomes nearly flat, at least for $M_{200} < 10^{13} \, \mathrm{M_\odot}$.
The SHM relation still drops at higher halo masses, in small part because
our model variations do not affect star formation within satellite galaxies
(see Fig.~\ref{cent_sat_frac}), and in larger part because massive haloes
are less relaxed and therefore contain a larger mass fraction within satellite 
subhaloes (and because these satellites are better resolved in more massive 
haloes). 
This is usually of secondary importance for galaxy stellar masses, 
since low-mass satellite galaxies ($M_{\star} \ll 10^{10} \, \mathrm{M_\odot}$)
are extremely inefficient at forming stars, meaning that satellites only
contribute significantly to the total stellar mass within $R_{200}$
for haloes that are massive enough to in turn host very massive satellite 
galaxies, as seen in Fig.~\ref{shm_eagle}.
For the model considered here, if all of the  N12 model terms are fixed, 
then low-mass satellites are dramatically more efficient in forming stars,
such that the higher mass fraction in sub-structures found in group
and cluster-mass haloes is instead the main reason for why there is relatively
less stellar mass in the central subhalo. This point is further explained in Appendix~\ref{ap_dm_first}.

Relative to this baseline scenario (and relative to our fiducial implementation of the N12 model), 
the mass dependencies of the outflow terms (red) are clearly the most important for shaping the low-mass
slope of the SHM relation, but still only account for about one half of
the slope in this regime, with the first-time accretion (blue) and recycling (green)
terms clearly all playing a role. The high-mass slope of the SHM relation 
is set mostly by the mass dependence of the first-time accretion terms 
(specifically gas accretion onto the ISM, $G_{\mathrm{acc}}^{\mathrm{gal}}$),
with the outflow terms playing a secondary role. Newly in this panel, we also show the effect
of fixing the star formation efficiency term ($G_{\mathrm{SF}}$, purple), which
has no discernible effect. 

Finally, if we compare to the simplified picture set out in Eqn.~\ref{dave_eq} 
and Fig.~\ref{dave_fig} of Section~\ref{dave_sec} (note that a flat line
in Fig.~\ref{dave_fig} means the corresponding process does not affect
the shape of the SHM relation, whereas the opposite is true for a flat
line in Fig.~\ref{m12_fix_fig}), we see that the full N12 model has more or less borne out our basic expectations.
As shown in Fig.~\ref{accuracy_fig}, the full model is much more successful at reproducing the \eagle
SHM relation, which we expect is because it accounts for the full redshift evolution of different
processes at a given halo mass, and also accounts for the effects of galaxy and 
subhalo mergers\footnote{It should also be noted that the full N12 model contains more
model terms at a given redshift than the simplified model discussed
in Section~\ref{dave_sec}, but we expect this to be of secondary importance.
Indeed, \cite{Neistein12} show that simplifying their fiducial model by
reducing the number of tracked gas reservoirs (thereby also reducing the number of
model terms) does not comprimise the accuracy with which the model reproduces
the base simulation.}. Compared to Fig.~\ref{dave_fig}, the full N12 model shows
that galaxy-scale recycling is indeed a secondary (but not negligible) effect for
the low-mass slope of the SHM relation. At higher masses, Fig.~\ref{dave_fig}
implies that galaxy-scale recycling should also be mildly important, which 
is not the case in the full N12 model. This is (in part) because in deriving Eqn.~\ref{dave_eq}
we made the approximation that $\eta^{\mathrm{gal}} \gg 1$, which is not
the case for high-mass galaxies (see Fig.~\ref{coeff_fig}).
Relative to Eqn.~\ref{dave_eq}, the full N12 model separates out the ``preventative
feedback'' term ($f_{\mathrm{prev}}$) into halo-scale accretion ($f_{\mathrm{acc}}^{\mathrm{halo}}$),
outflows ($\eta^{\mathrm{halo}}$) and recycling ($G_{\mathrm{ret}}^{\mathrm{halo}}$), and
also galaxy-scale accretion ($G_{\mathrm{acc}}^{\mathrm{gal}}$). With these terms
fully separated, we can now appreciate that it is the halo mass dependence of the 
halo-scale outflows that provides the most important
contribution to $f_{\mathrm{prev}}$ for the low-mass SHM slope, and that the mass dependence of the first-time 
galaxy-scale infall efficiency term ($G_{\mathrm{acc}}^{\mathrm{gal}}$) is most important 
for the high-mass SHM slope.

\subsection{What sets the redshift evolution of the SHM relation?}

\begin{figure*}
\begin{center}
\includegraphics[width=40pc]{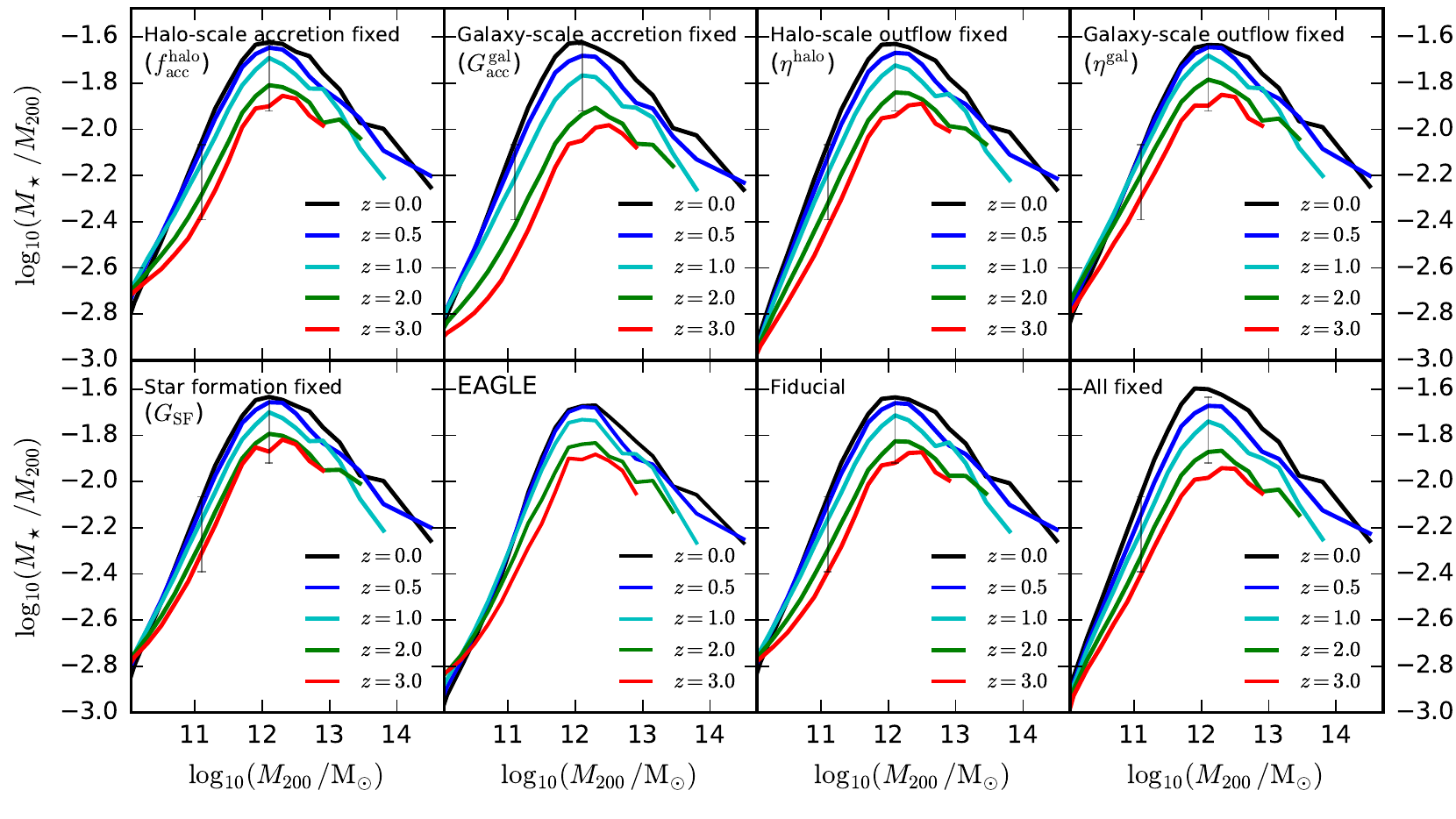}
\caption{The impact of removing the redshift dependence of one (or more) term(s) 
in the N12 model by fixing its value(s) to the one(s) corresponding to $z=1$.
Note that if we hold the terms related to timescales 
($G_{\mathrm{acc}}^{\mathrm{gal}}$, $G_{\mathrm{SF}}$, 
$G_{\mathrm{ret}}^{\mathrm{gal}}$, $G_{\mathrm{ret}}^{\mathrm{halo}}$)
constant at fixed halo mass, then the associated timescales 
still scale with the Hubble time.
The median ratio of galaxy stellar mass to halo mass is plotted as a function of halo mass
for central galaxies. 
Different line colours correspond to different redshifts.
The lower-middle-left panel shows the evolution of the SHM relation in \eagle, and the lower-middle-right
panel shows the corresponding evolution in the N12 model.
The lower-far-right (``All'') panel shows the model when each of the parameters $f_{\mathrm{acc}}^{\mathrm{halo}}$, 
$G_{\mathrm{acc}}^{\mathrm{gal}}$, $G_{\mathrm{SF}}$, $\eta^{\mathrm{gal}}$, $\eta^{\mathrm{halo}}$, 
$G_{\mathrm{ret}}^{\mathrm{gal}}$, $G_{\mathrm{ret}}^{\mathrm{halo}}$ are held fixed to their values at $z=1$.
Other panels show the model when only one of these terms is fixed to its value at $z=1$.
To save space, we do not show panels for the recycling terms as they are indistinguishable from the fiducial model.
As a reference point, the grey error bars show the evolution of the SHM relation for the fiducial N12 model
(lower-middle-right) at $M_{200} = 10^{11}$ and $10^{12} \, \mathrm{M_\odot}$, over the range $0<z<3$.
Most of the evolution of the SHM relation in \eagle seems to be set by the evolution of the star
formation efficiency ($G_{\mathrm{SF}}$), galaxy-scale gas accretion ($G_{\mathrm{acc}}^{\mathrm{gal}}$, i.e., the
efficiency of first-time infall from the CGM to the ISM), galaxy-scale outflows ($\eta^{\mathrm{gal}}$), 
and halo-scale outflows ($\eta^{\mathrm{halo}}$).
}
\label{z_fix_fig}
\end{center}
\end{figure*}

To explore how (and which) physical processes in \eagle shape the evolution of the
SHM relation, we fix a single term in the N12 model to its value
at $z=1$ (but keeping the halo mass dependence), thereby removing any redshift dependence for that
term. Note that for the terms that are related to timescales 
($G_{\mathrm{acc}}^{\mathrm{gal}}$, $G_{\mathrm{SF}}$, $G_{\mathrm{ret}}^{\mathrm{gal}}$, 
$G_{\mathrm{ret}}^{\mathrm{halo}}$), the associated timescale is therefore still
assumed to scale with the Hubble time. For example, the 
star formation efficiency within the ISM 
($G_{\mathrm{SF}} \equiv t \, \dot{M}_\star \, / M_{\mathrm{ISM}} \equiv t/\tau_{\mathrm{SF}}$)
is fixed such that the ISM depletion time $\tau_{\mathrm{SF}}$ scales with the Hubble time, 
and so is shorter at higher redshift.

Fig.~\ref{z_fix_fig} shows the results of this exercise, comparing \eagle
(lower-middle-left panel) to different variations of the N12 model. The fiducial
model (being approximate in nature) slightly over-predicts the evolution at
fixed halo mass compared to \eagle, particularly for 
$M_{200} \sim 10^{11} \, \mathrm{M_\odot}$, 
but is nonetheless a reasonable representation of the underlying simulation.
In the base \eagle simulation, stellar masses increase at fixed halo mass
with decreasing redshift, at least for $z<3$ and for $M_{200} \gtrsim 10^{11} \, \mathrm{M_\odot}$,
with the clearest redshift evolution seen for $M_{200} \sim 10^{12} \, \mathrm{M_\odot}$.
This could imply that low-redshift haloes are more efficient at forming stars
at fixed halo mass, but we will show presently that this interpretation is 
flawed.

Stepping through the various model terms in turn, we see that fixing
either $G_{\mathrm{SF}}$ (star formation efficiency within the ISM) or 
$\eta^{\mathrm{gal}}$ (galaxy-scale outflows) reduces the level of SHM evolution.
Conversely, fixing either $G_{\mathrm{acc}}^{\mathrm{gal}}$ (first-time infall
from the CGM to ISM) or $\eta^{\mathrm{halo}}$ (halo-scale outflows) increases
the level of SHM evolution. 
This can be visually appreciated by comparing to the grey error bars, which show
the range in evolution for different halo masses in the fiducial model.
The interpretation of these trends is straightforward, as visual inspection
of Fig.~\ref{coeff_fig} shows that these four model terms evolve at fixed
halo mass in a manner that is consistent with the trends seen in Fig.~\ref{z_fix_fig}.
Halo-scale outflows become more efficient at lower redshift, but galaxy-scale
outflows are more efficient at higher redshift. Star formation in the ISM
is more efficient (relative to the Hubble time) at lower redshift, whereas
first-time gas accretion onto galaxies is more efficient at higher redshift.
We note from Fig.~\ref{coeff_fig} that the efficiencies of galaxy and
halo-scale gas recycling also vary comparably with redshift (and in opposite
directions relative to each other), but we find that fixing these terms
has negligible effect on the evolution of the SHM relation (probably because
of their relatively weak impact, see Fig.~\ref{boost_fig}), 
and so we omit them from Fig.~\ref{z_fix_fig} to save space.

\begin{figure*}
\begin{center}
\includegraphics[width=40pc]{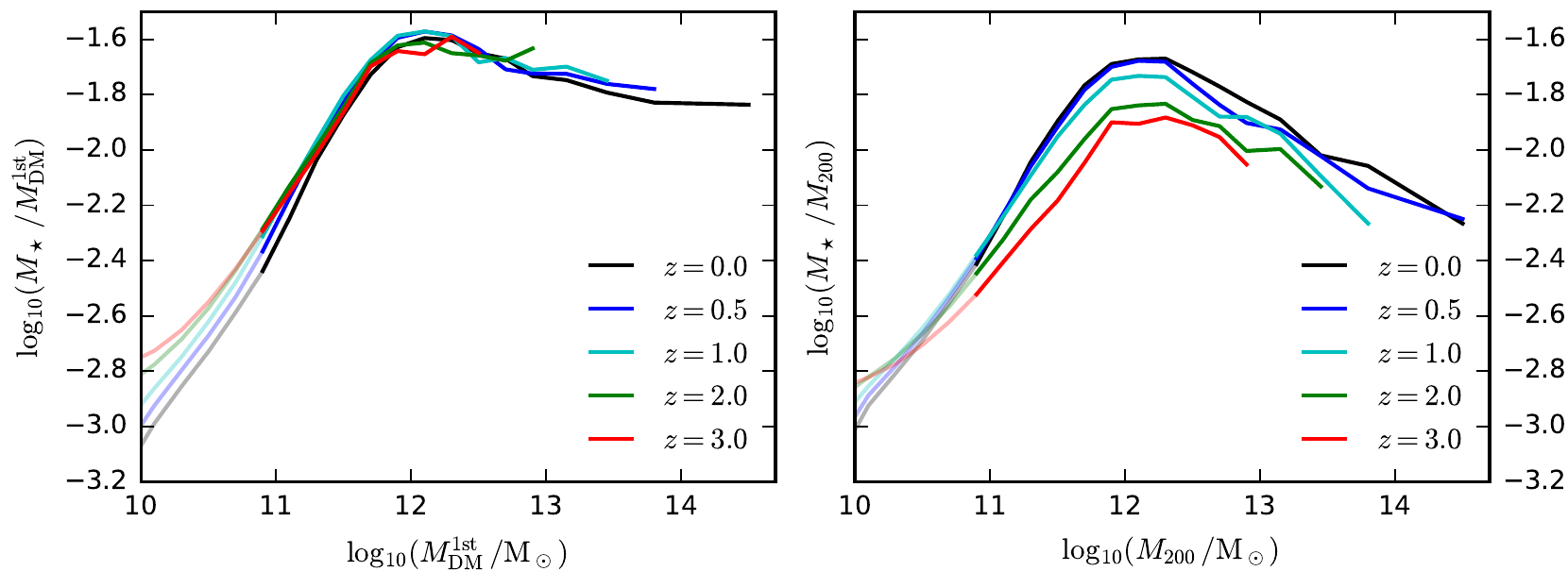}
\caption{\textit{Left:} the evolution of the \eagle SHM relation for central galaxies, but replacing $M_{200}$ 
with $M_{\mathrm{DM}}^{\mathrm{1st}}$, the cumulative mass of
dark matter accreted for the first time onto progenitors of the central
subhalo.
\textit{Right:} the same evolution of the SHM relation but using $M_{200}$ as the halo mass variable,
to act as a visual reference.
For each line, lighter shades indicate the halo mass range for which galaxies contain on average fewer than $100$ star particles.
Compared to the conventional SHM definition shown in the right panel,
the ratio of $M_\star / M_{\mathrm{DM}}^{\mathrm{1st}}$ exhibits almost
no redshift evolution at fixed halo mass for $M_{DM}^{\mathrm{1st}} > 10^{11} \, \mathrm{M_\odot}$.
In general, it is expected that $M_\star$ should trace $M_{\mathrm{DM}}^{\mathrm{1st}}$
more closely than $M_{200}$ (see main text). It follows therefore 
that there is overall very little evolution
in the efficiency of galaxy formation with cosmic time in \eagle, given the amount of expected
baryonic accretion onto a halo of a given mass.
}
\label{shm_dm_first_acc}
\end{center}
\end{figure*}

Interestingly, these various effects conspire to nearly cancel out for
the model variant where we simultaneously fix all of the model terms
to their values at $z=1$ (``All'', lower-far-right panel), with this
variant exhibiting slightly more redshift evolution than the fiducial model.
This is at first glance puzzling, since if the efficiency of all
terms are constant with redshift, there is no reason for there
to be any evolution in the SHM relation. The explanation for this
paradox is shown in the left panel of Fig.~\ref{shm_dm_first_acc}, which shows the
evolution of the \eagle SHM relation, but replaces $M_{200}$ 
with $M_{\mathrm{DM}}^{\mathrm{1st}}$, the cumulative mass of
dark matter accreted (for the first time) onto progenitors of the central
subhalo. As a reminder, the time derivative of this quantity acts
as the source term for the model; i.e., we assume that gas
being accreted for the first time at the virial radius traces
the rate with which dark matter is being accreted for the first time.

Fig.~\ref{shm_dm_first_acc} shows that when compared to the conventional
SHM definition using $M_{200}$ (right panel), the SHM relation
defined instead with $M_{\mathrm{DM}}^{\mathrm{1st}}$ (left panel) exhibits
almost no redshift evolution at fixed halo mass, aside from
for low-mass haloes that are poorly resolved. $M_{200}$ differs from
$M_{\mathrm{DM}}^{\mathrm{1st}}$ in several important respects.
First, $M_{200}$ includes the mass within satellite subhaloes, whereas
we show $M_{\mathrm{DM}}^{\mathrm{1st}}$ for the central subhalo only
(note that, following convention, $M_\star$ is also shown for the central
subhalo only). At fixed host halo mass, the fraction of total mass in satellite subhaloes increases
with increasing redshift in \eagle. In addition, $M_{200}$ also includes
the total mass that has been stripped from satellite subhaloes ($M_{\mathrm{DM}}^{\mathrm{1st}}$
does not). In general, we expect galaxy
stellar masses to better trace $M_{\mathrm{DM}}^{\mathrm{1st}}$ than
the instantaneous mass of a subhalo after stripping, since dark matter
is much more readily stripped from a satellite subhalo compared
to stars.

A second factor is that significant amounts of accreted dark matter will
move back out beyond $R_{200}$, defining the so-called ``splashback'' radius
at larger spatial scales \cite[e.g.,][]{Diemer17b}. Some of this ejected dark matter will then
be re-accreted at later times, actually comprising the majority of
the total dark matter accretion at late times \cite[e.g.,][]{Wright20}.
$M_{\mathrm{DM}}^{\mathrm{1st}}$ is by definition insensitive to this process, but
$M_{200}$ is affected.
As a third and final consideration, $M_{200}$ includes the contribution of baryons, and the baryon
fraction evolves in \eagle at fixed halo mass.

Putting these effects together\footnote{We have performed a
preliminary investigation of which effect is more important, and find
simply that they likely all play a role.}, 
it is possible for the SHM relation
to evolve even if the galaxy formation efficiency were to remain constant with
time, due to the evolution of $M_{200}$ for a given halo. Stars
do form continuously in the \eagle simulation, but the evolution
of the SHM relation is mostly set by the evolution of $M_{200}$,
and not by an evolving efficiency with which haloes convert gas
into stars. While difficult to test observationally,
the same may well hold for the SHM relation in the real Universe, when 
using the conventional halo mass definition of $M_{200}$ 
(or similar definitions). We note that \cite{Behroozi19}
use observations to infer similar evolution of the SHM relation to \eagle, in that
they find the normalization increases as a function of cosmic
time for halo masses $\sim 10^{12} \, \mathrm{M_\odot}$ until
$z=1$, and is approximately constant afterwards for $0<z<1$.
We have verified that the same is true in \eagle if we
use their halo mass definition \cite[which is taken from ][]{Bryan98}.

\section{What sets the relationship between the mass of the ISM, CGM, and halo mass?}
\label{ism_cgm_sec}

\begin{figure*}
\begin{center}
\includegraphics[width=40pc]{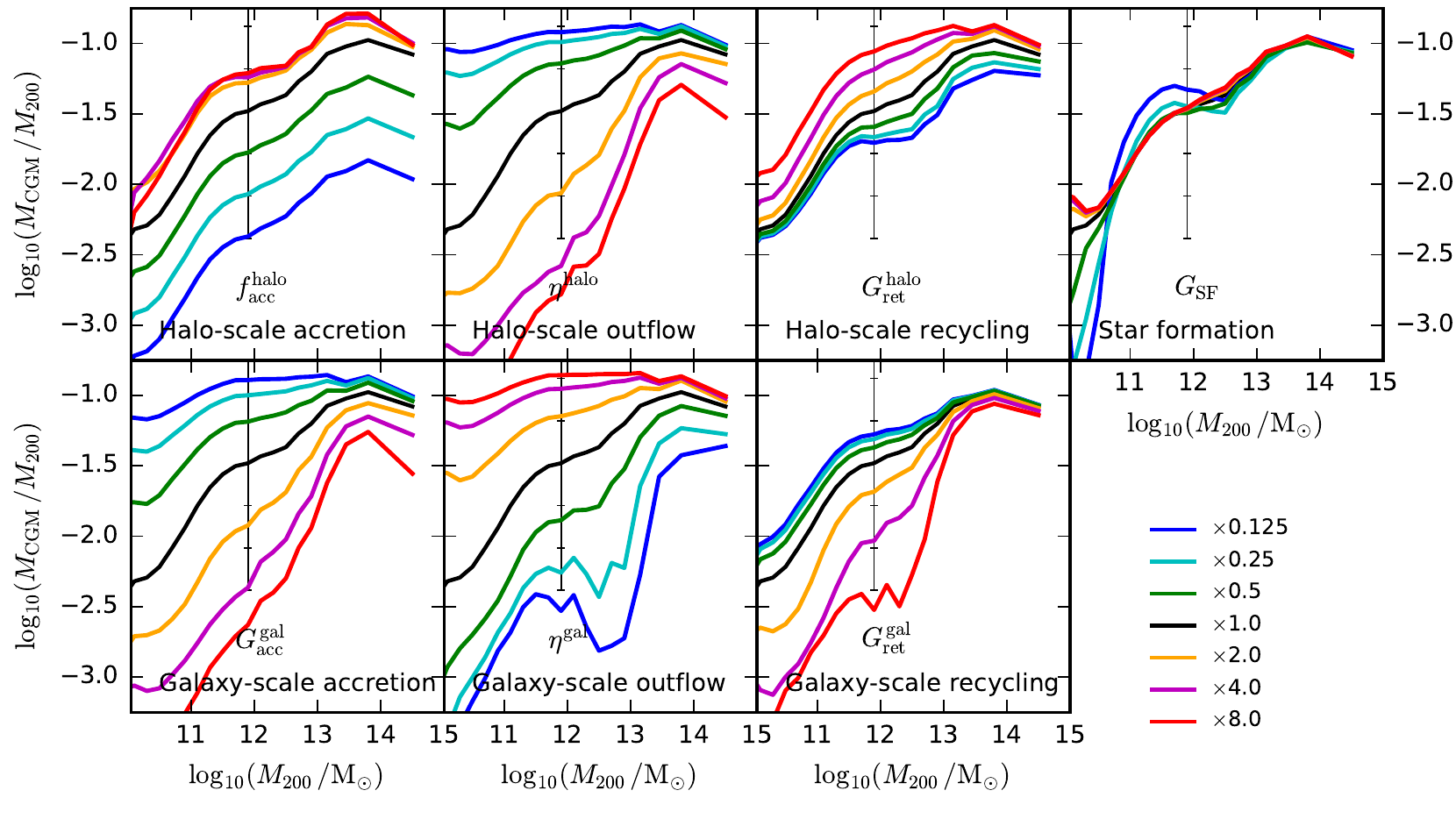}
\caption{The impact on the mass of the CGM of multiplying individual terms in our model 
(see \protect Eqn.~\ref{ODE_MATRIX}) by a constant. Masses are plotted for central galaxies at $z=0$.
The figure format follows \protect Fig.~\ref{boost_fig}, replacing
galaxy stellar mass with CGM mass on the y-axis.
Compared to \protect Fig.~\ref{boost_fig}, it is apparent that the 
CGM is more sensitive to a wider range of the model terms 
(e.g. galactic-scale recycling).}
\label{cgm_boost_fig}
\end{center}
\end{figure*}

\begin{figure*}
\begin{center}
\includegraphics[width=40pc]{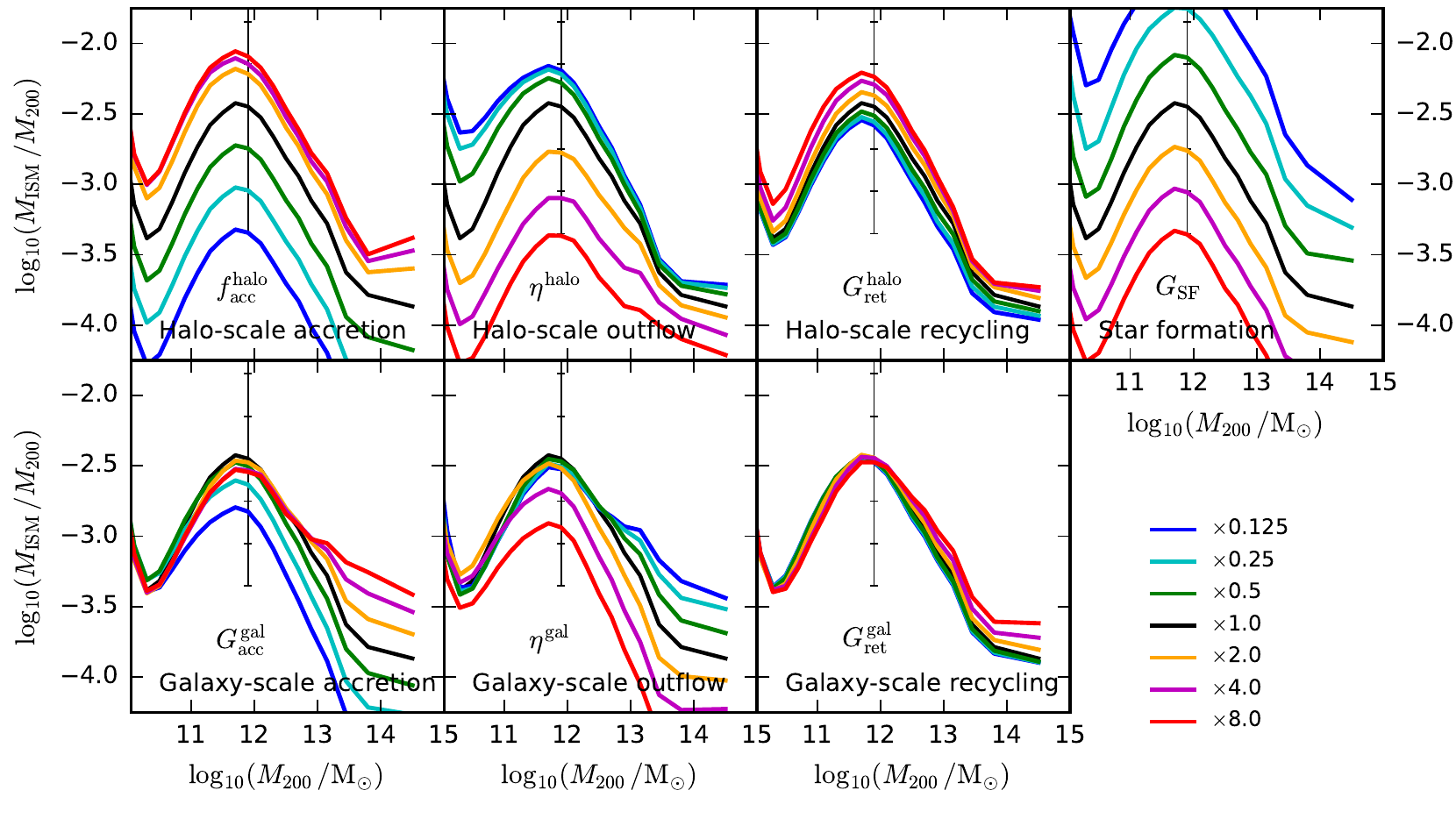}
\caption{The same as \protect Fig.~\ref{cgm_boost_fig}, but in this case showing the ISM mass
(again at $z=0$).
The ISM (unlike stars and the CGM) is very sensitive to the star formation efficiency term
($G_{\mathrm{SF}}$, upper-far-right panel).
The mass in the ISM is also slightly more sensitive to halo-scale outflows and recycling 
than the mass in stars is, but is slightly less sensitive to galaxy-scale outflows and recycling.
This is because the ISM mass traces recent gas accretion and outflow activity, whereas 
galaxy stellar masses reflect the full time-integrated activity. 
Note that the upturn in ISM-mass for very low-mass haloes 
($M_{200} < 10^{10.5} \, \mathrm{M_\odot}$) is connected to our adopted definition of the
ISM: specifically the inclusion of low-metallicity dense gas that does not pass
the metallicity-dependent SF threshold in \eagle (see Section~\ref{simul_sec}).
}
\label{ism_boost_fig}
\end{center}
\end{figure*}

While our primary focus has been the SHM relation, we now briefly extend our analysis
to consider how galactic star formation, inflows, and outflows affect the relationship
between halo mass and the masses of the ISM and CGM. 
Fig.~\ref{cgm_boost_fig} and Fig.~\ref{ism_boost_fig} show respectively how 
the CGM and ISM masses at $z=0$ respond to multiplying individual terms in our
model by a constant. These figures are analogous to Fig.~\ref{boost_fig},
which shows the corresponding information for galaxy stellar masses.

Starting with the CGM in Fig.~\ref{cgm_boost_fig}, it is apparent that the CGM 
is in general more sensitive to the chosen model parameters, when compared
to the mass in stars (Fig.~\ref{boost_fig}). This is true for all of the model terms except for
the first time halo-scale gas accretion term ($f_{\mathrm{acc}}^{\mathrm{halo}}$, 
upper-far-left panel), since this term is directly proportional to the 
source term in Eqn.~\ref{ODE_MATRIX}. In some cases,
the mass in the CGM even responds super-linearly to changes in model
parameters: this is the case for halo-scale outflows ($\eta^{\mathrm{halo}}$,
upper-middle-left panel), for first time galaxy-scale accretion 
($G_{\mathrm{acc}}^{\mathrm{gal}}$, lower-far-left panel), and for 
galaxy-scale outflows ($\eta^{\mathrm{gal}}$, lower-middle panel).

It is generally accepted that the CGM should be a sensitive tracer of gas flows and
feedback processes, since many of the metals produced in stars are believed to
be ejected from galaxies \cite[e.g.,][]{Peeples14}, and because galaxy properties
(ISM and stars) are often degenerate with respect to competing scenarios
for galactic inflows, outflows and recycling \cite[e.g.,][]{Mitchell14, Mitchell20a, Pandya20}.
Fig.~\ref{cgm_boost_fig} bears out this expectation. Comparing
Fig.~\ref{cgm_boost_fig} with Fig.~\ref{boost_fig}, it is particularly
noteworthy that the mass in the CGM is much more sensitive to both 
galaxy- and halo-scale recycling than the mass in stars is.
While halo-scale recycling boosts the CGM mass at all halo masses, 
galaxy-scale recycling suppresses the CGM for $M_{200} < 10^{13} \, \mathrm{M_\odot}$
and has little effect for higher halo masses.
Note that our model only predicts the total mass of the CGM,
and does provide information about the relative mass in different
phases. It is possible (for example) that the cool/dense observable
phases of the CGM are less sensitive than the CGM as a whole.

Considering instead the mass of the ISM in Fig.~\ref{ism_boost_fig},
the most obvious difference is that the ISM mass depends negatively
and approximately linearly with the value of the star formation
efficiency term ($G_{\mathrm{SF}} \equiv \dot{M}_\star \, t \, / M_{\mathrm{ISM}}$, 
upper-far-right panel). In contrast, the mass in stars and in the
CGM depends only very weakly on this parameter. This is a well
known result that galaxy formation is self-regulated by feedback,
in the sense that star formation rates and outflow rates (and
therefore the masses of stars and the CGM) respond
such that they roughly balance the galaxy-scale accretion rate
\cite[e.g.,][]{Finlator08,Schaye10,Dave12,Lilly13,Sharma20}.
This renders $G_{\mathrm{SF}}$ inconsequential for star formation
rates and outflow rates unless $G_{\mathrm{SF}} < 1$ (at which
point gas consumption in the ISM is an important bottleneck).
This self-regulation is achieved as a result of the ISM
mass increasing or decreasing in response to the balance
between accretion and the efficiencies of outflows and star
formation. As such, a shorter gas consumption time scale requires less ISM to produce 
the star formation rate that is required for outflows plus 
star formation to balance inflows, meaning that 
the ISM mass is very sensitive to the star formation efficiency
\cite[][]{Schaye10,Haas13}.

Comparing the ISM mass shown in Fig.~\ref{ism_boost_fig} to the stellar
mass shown in Fig.~\ref{boost_fig}, it is also notable that the
ISM at $z=0$ is more sensitive to halo-scale outflows
and recycling ($\eta^{\mathrm{halo}}$, $G_{\mathrm{ret}}^{\mathrm{halo}}$), 
whereas the stellar mass is comparatively more
sensitive to galaxy-scale outflows and recycling 
($\eta^{\mathrm{gal}}$, $G_{\mathrm{ret}}^{\mathrm{gal}}$). 
Gas consumption times in the ISM are generally
shorter than the age of the Universe in \eagle (i.e., $G_{\mathrm{SF}} > 1$,
see Fig.~\ref{coeff_fig}), and the effective consumption time is
even shorter once galaxy-scale outflows are accounted for.
This means that the ISM at a given redshift is only sensitive
to recent accretion, star formation, and outflow activity. In contrast,
galaxy stellar masses reflect the integrated activity over
the entire history of a galaxy. From Fig.~\ref{coeff_fig},
we see that halo-scale outflows are more efficient at low
redshift in \eagle, whereas the efficiency of galaxy-scale outflows 
increases with increasing redshift. This means in turn that the
$z=0$ ISM mass is more sensitive to $\eta^{\mathrm{halo}}$,
whereas the $z=0$ stellar mass is (comparatively) more
sensitive to $\eta^{\mathrm{gal}}$, as many of the
stars in galaxies at $z=0$ were formed for $z>1$.

\section{Summary}
\label{summary_sec}

The median SHM relation, $\frac{M_\star}{M_{200}}(M_{200},z)$,
is a fundamental diagnostic of the efficiency of galaxy formation within
the context of the working cosmological model.
We have introduced a method to explain how the SHM
relation predicted by hydrodynamical simulations is shaped,
disentangling the relative contributions
of gas consumption by star formation, first-time inflows 
(i.e. accretion of gas that has not been
ejected in the past), gas ejection due to outflows, and wind recycling.
We used measurements of gas flows in the \eagle simulations
from \cite{Mitchell20b,Mitchell20a}, averaging as a function
of halo mass and redshift. These measurements
then provide coefficients in a set of coupled linear
differential equations for the evolution of the mass
fractions in stars, ISM, CGM, and gas that is ejected
from the halo, with the first time gas accretion
at the virial radius acting as the source term.
By integrating these equations along the merger tree
of each individual galaxy in the \eagle simulation,
we constructed a model that then computes the mass fractions
for each individual \eagle galaxy.
By inspection of the model coefficients, and by modifying the coefficients
by hand, we built an understanding of
what shapes the SHM relation and its evolution with redshift.

When modifying individual model terms by a constant
multiplicative factor, we find that the normalization
of the SHM relation
is most sensitive to the efficiencies of first-time gas
accretion and ejection by outflows, and is less sensitive
to the efficiency of wind recycling, and of gas
consumption by star formation (Fig.~\ref{boost_fig}).
Cosmological first-time gas accretion at the halo virial
radius sets the main boundary condition (the maximum possible
value of $M_\star \, / M_{200}$). For a fixed first-time
halo accretion rate, $M_\star$ is suppressed by
galaxy-scale outflows (for all masses), halo-scale
outflows (mostly for $M_{200} < 10^{12} \, \mathrm{M_\odot}$),
and by the finite efficiency of galaxy-scale gas
accretion (including the effect of preventative
feedback, mostly for $M_{200} > 10^{12} \, \mathrm{M_\odot}$).
For a fixed first-time halo accretion rate, $M_\star$
is increased by gas recycling, although the effect
is smaller than that of outflows.
The efficiency of gas consumption in the ISM via
star formation would have to be reduced dramatically
for it to become a bottleneck, and therefore has little
effect on the $z=0$ SHM relation in \eagle.

Our first main result is that the shape of 
the SHM relation is driven primarily by the ejection of
gas via outflows for
$M_{200} < 10^{12} \, \mathrm{M_\odot}$, and by the 
(in)efficiency of first-time gas infall from the CGM to the
ISM in more massive haloes (Fig.~\ref{m12_fix_fig}), although
the drop in $M_\star/M_{200}$ for $M_{200} > 10^{12} \, \mathrm{M_\odot}$
is actually driven mostly by the conventional choice
to exclude satellite galaxies and the diffuse stellar halo
from the stellar mass used in SHM relation (Fig.~\ref{shm_eagle}). 
For $M_{200} < 10^{12} \, \mathrm{M_\odot}$, the steep
$M_\star \, /M_{200} \propto M_{200}$ slope of the SHM relation
is shaped by the ejection of gas via outflows at both the galaxy 
(i.e., gas ejected from the ISM) and halo (i.e., gas ejected beyond the virial radius)
scales. Interestingly, including the halo mass dependence of either
the galaxy or halo-scale gas ejection terms in isolation is sufficient
to explain most of the total effect of ejection on the SHM shape. 
Including the halo mass dependencies of both galaxy- and halo-scale
outflow rates together has only a weak additional effect,
since more efficient galaxy-scale gas ejection reduces
the gas ejection rate at the virial radius (due to reduced
star formation rates), and vice versa.
In addition, for $M_{200} < 10^{12} \, \mathrm{M_\odot}$
secondary roles are played by preventative
feedback (i.e., a reduction in the first-time inflow
of gas) at the scale of the halo virial radius,
and by halo-scale recycling of ejected gas.

Our second main result is that the redshift evolution of
the SHM relation (which drops with increasing
redshift at fixed halo mass for $z>1$ in \eagle)
is driven mostly by the evolution of halo properties,
rather than any change in the combined net efficiency
of inflows, star formation, outflows, and recycling.
Specifically, the baryon fraction within $R_{200}$ (i.e., backsplash),
the mass fraction in satellite subhaloes (which
contribute to $M_{200}$ but not to $M_\star$, by
convention), the ejection and re-accretion of
dark matter at $R_{200}$, and satellite subhalo
mass loss (baryons plus dark matter) all evolve
with redshift at fixed halo mass, driving in turn
the redshift evolution of the SHM relation 
(Fig.~\ref{shm_dm_first_acc}).
In contrast, when, instead of $M_{200}$, we use the 
cumulative mass of dark matter that is accreted for the first time, 
then the SHM relation does not evolve.

Finally, we also briefly examined how star formation
and gas flows affect the relationship between halo
mass, and the masses of both the ISM and the CGM.
Of the three baryonic components inside
haloes (stars, ISM, CGM), the mass of the CGM is generally
the most sensitive to the effects
of inflows and outflows (Fig.~\ref{cgm_boost_fig}).
Notably, gas recycling (at both the halo
and galaxy scales) has a much larger impact on the CGM
than on the stars (Fig.~\ref{boost_fig}) or
on the ISM (Fig.~\ref{ism_boost_fig}).
The ISM is the only reservoir that is sensitive
to the ISM gas consumption efficiency via 
star formation (Fig.~\ref{ism_boost_fig}), 
reflecting the well-known result 
that galaxies self-regulate by increasing
or decreasing star formation and outflow
rates to balance changes in the galaxy-scale
gas accretion rate. Compared to the mass in
stars, the ISM mass at $z=0$ is more sensitive
to halo-scale gas flows (Fig.~\ref{ism_boost_fig}), whereas the 
stellar mass is comparatively more sensitive
to galaxy-scale gas flows (Fig.~\ref{boost_fig}). This is because
the ISM traces only recent accretion, star formation
and outflow activity, combined with the fact that galaxy-scale
outflows are more efficient at high redshift in \eagle,
whereas halo-scale outflows are more efficient
at low redshift (Fig.~\ref{coeff_fig}).

In conclusion, we have estimated the impact of different types of gas
flows on the SHM relation and on the mass fractions of the ISM and CGM
in the \eagle hydrodynamical simulation. By varying the values of
individual terms of a model that by design reproduces the
gas flows measured for \eagle, we investigated the effects of varying the
efficiency of star formation in the ISM and the rates of, respectively,
the first time gas accretion, outflows and recycling, where we
distinguished gas flows on galaxy and halo scales. While cosmological
first-time accretion onto the halo sets the overall amplitude of the SHM
relation, its shape is modified mainly by outflows, with halo-scale
outflows being relatively more important at lower masses. The outflows
do not only have a direct impact on the stellar mass, they also affect
it indirectly by suppressing gas accretion onto galaxies, especially for
more massive objects. Recycling, either on galaxy or halo scales, has
only a minor effect on the SHM relation in \eagle, though it does have a
large impact on the CGM. We also found that the evolution of the SHM
relation nearly vanishes when we plot the ratio of the stellar mass to
the cumulative mass of dark matter that fell in for the first time, as
opposed to plotting the ration with the total halo mass. This suggests
that the evolution of the SHM relation is set by the definitions of the
galaxy and halo, rather than by evolution in the true efficiency of
galaxy formation.

It would be of considerable interest to apply our methodology to another
state-of-the-art hydrodynamical simulation that provides a good match to
the SHM relation inferred from observations. This would reveal how
generic our conclusions are.

\section*{Acknowledgements}

We thank Eyal Neistein for sharing the source code used
for the original implementation of his modelling technique,
and for directly inspiring the methodology we use in this paper. 
We would also like to thank John Helly for producing and sharing the halo
merger trees that form the backbone of our analysis.

This work used the DiRAC@Durham facility managed by the Institute for
Computational Cosmology on behalf of the STFC DiRAC HPC Facility
(www.dirac.ac.uk). The equipment was funded by BEIS capital funding
via STFC capital grants ST/K00042X/1, ST/P002293/1, ST/R002371/1 and
ST/S002502/1, Durham University and STFC operations grant
ST/R000832/1. DiRAC is part of the National e-Infrastructure.

This work was supported by Vici grant 639.043.409 from
the Netherlands Organisation for Scientific Research (NWO).

\section*{Data availability}

The data underlying this article will be shared on reasonable request to the corresponding author.
Raw particle data and merger trees for the \eagle simulations have been publicly released \cite[][]{McAlpine16}.

\bibliographystyle{mn2e}
\bibliography{bibliography}

\begin{thebibliography}{82}
\expandafter\ifx\csname natexlab\endcsname\relax\def\natexlab#1{#1}\fi

\bibitem[{{Angl{\'e}s-Alc{\'a}zar}
  {et~al}\mbox{.}(2017){Angl{\'e}s-Alc{\'a}zar}, {Faucher-Gigu{\`e}re},
  {Kere{\v s}}, {Hopkins}, {Quataert}, \& {Murray}}]{AnglesAlcazar17}
{Angl{\'e}s-Alc{\'a}zar} D., {Faucher-Gigu{\`e}re} C.-A., {Kere{\v s}} D.,
  {Hopkins} P.~F., {Quataert} E., {Murray} N., 2017, \mnras, 470, 4698

\bibitem[{{Bauermeister} {et~al}\mbox{.}(2010){Bauermeister}, {Blitz}, \&
  {Ma}}]{Bauermeister10}
{Bauermeister} A., {Blitz} L., {Ma} C.-P., 2010, \apj, 717, 323

\bibitem[{{Behroozi} {et~al}\mbox{.}(2019){Behroozi}, {Wechsler}, {Hearin}, \&
  {Conroy}}]{Behroozi19}
{Behroozi} P., {Wechsler} R.~H., {Hearin} A.~P., {Conroy} C., 2019, \mnras,
  488, 3143

\bibitem[{{Behroozi} {et~al}\mbox{.}(2010){Behroozi}, {Conroy}, \&
  {Wechsler}}]{Behroozi10}
{Behroozi} P.~S., {Conroy} C., {Wechsler} R.~H., 2010, \apj, 717, 379

\bibitem[{{Bell} {et~al}\mbox{.}(2003){Bell}, {McIntosh}, {Katz}, \&
  {Weinberg}}]{Bell03}
{Bell} E.~F., {McIntosh} D.~H., {Katz} N., {Weinberg} M.~D., 2003, \apjs, 149,
  289

\bibitem[{{Berlind} \& {Weinberg}(2002)}]{Berlind02}
{Berlind} A.~A., {Weinberg} D.~H., 2002, \apj, 575, 587

\bibitem[{{Booth} \& {Schaye}(2009)}]{Booth09}
{Booth} C.~M., {Schaye} J., 2009, \mnras, 398, 53

\bibitem[{{Borrow} {et~al}\mbox{.}(2020){Borrow}, {Angl{\'e}s-Alc{\'a}zar}, \&
  {Dav{\'e}}}]{Borrow20}
{Borrow} J., {Angl{\'e}s-Alc{\'a}zar} D., {Dav{\'e}} R., 2020, \mnras, 491,
  6102

\bibitem[{{Bower} {et~al}\mbox{.}(2006){Bower}, {Benson}, {Malbon}, {Helly},
  {Frenk}, {Baugh}, {Cole}, \& {Lacey}}]{Bower06}
{Bower} R.~G., {Benson} A.~J., {Malbon} R., {Helly} J.~C., {Frenk} C.~S.,
  {Baugh} C.~M., {Cole} S., {Lacey} C.~G., 2006, \mnras, 370, 645

\bibitem[{{Bryan} \& {Norman}(1998)}]{Bryan98}
{Bryan} G.~L., {Norman} M.~L., 1998, \apj, 495, 80

\bibitem[{{Clauwens} {et~al}\mbox{.}(2018){Clauwens}, {Schaye}, {Franx}, \&
  {Bower}}]{Clauwens18}
{Clauwens} B., {Schaye} J., {Franx} M., {Bower} R.~G., 2018, \mnras, 478, 3994

\bibitem[{{Cole}(1991)}]{Cole91}
{Cole} S., 1991, \apj, 367, 45

\bibitem[{{Conroy} {et~al}\mbox{.}(2006){Conroy}, {Wechsler}, \&
  {Kravtsov}}]{Conroy06}
{Conroy} C., {Wechsler} R.~H., {Kravtsov} A.~V., 2006, \apj, 647, 201

\bibitem[{{Correa} {et~al}\mbox{.}(2018){Correa}, {Schaye}, {van de Voort},
  {Duffy}, \& {Wyithe}}]{Correa18b}
{Correa} C.~A., {Schaye} J., {van de Voort} F., {Duffy} A.~R., {Wyithe} J.
  S.~B., 2018, \mnras, 478, 255

\bibitem[{{Crain} {et~al}\mbox{.}(2015){Crain}, {Schaye}, {Bower}, {Furlong},
  {Schaller}, {Theuns}, {Dalla Vecchia}, {Frenk}, {McCarthy}, {Helly},
  {Jenkins}, {Rosas-Guevara}, {White}, \& {Trayford}}]{Crain15}
{Crain} R.~A. {et~al.}, 2015, \mnras, 450, 1937

\bibitem[{{Croton} {et~al}\mbox{.}(2006){Croton}, {Springel}, {White}, {De
  Lucia}, {Frenk}, {Gao}, {Jenkins}, {Kauffmann}, {Navarro}, \&
  {Yoshida}}]{Croton06}
{Croton} D.~J. {et~al.}, 2006, \mnras, 365, 11

\bibitem[{{Dalla Vecchia} \& {Schaye}(2012)}]{DallaVecchia12}
{Dalla Vecchia} C., {Schaye} J., 2012, \mnras, 426, 140

\bibitem[{{Dav{\'e}} {et~al}\mbox{.}(2019){Dav{\'e}}, {Angl{\'e}s-Alc{\'a}zar},
  {Narayanan}, {Li}, {Rafieferantsoa}, \& {Appleby}}]{Dave19}
{Dav{\'e}} R., {Angl{\'e}s-Alc{\'a}zar} D., {Narayanan} D., {Li} Q.,
  {Rafieferantsoa} M.~H., {Appleby} S., 2019, \mnras

\bibitem[{{Dav{\'e}} {et~al}\mbox{.}(2012){Dav{\'e}}, {Finlator}, \&
  {Oppenheimer}}]{Dave12}
{Dav{\'e}} R., {Finlator} K., {Oppenheimer} B.~D., 2012, \mnras, 421, 98

\bibitem[{{De Lucia} \& {Blaizot}(2007)}]{deLucia07}
{De Lucia} G., {Blaizot} J., 2007, \mnras, 375, 2

\bibitem[{{Dekel} \& {Silk}(1986)}]{Dekel86}
{Dekel} A., {Silk} J., 1986, \apj, 303, 39

\bibitem[{{Diemer}(2017)}]{Diemer17b}
{Diemer} B., 2017, \apjs, 231, 5

\bibitem[{{Dolag} {et~al}\mbox{.}(2009){Dolag}, {Borgani}, {Murante}, \&
  {Springel}}]{Dolag09}
{Dolag} K., {Borgani} S., {Murante} G., {Springel} V., 2009, \mnras, 399, 497

\bibitem[{{Drory} {et~al}\mbox{.}(2005){Drory}, {Salvato}, {Gabasch}, {Bender},
  {Hopp}, {Feulner}, \& {Pannella}}]{Drory05}
{Drory} N., {Salvato} M., {Gabasch} A., {Bender} R., {Hopp} U., {Feulner} G.,
  {Pannella} M., 2005, \apjl, 619, L131

\bibitem[{{Dubois} {et~al}\mbox{.}(2014){Dubois}, {Pichon}, {Welker}, {Le
  Borgne}, {Devriendt}, {Laigle}, {Codis}, {Pogosyan}, {Arnouts}, {Benabed},
  {Bertin}, {Blaizot}, {Bouchet}, {Cardoso}, {Colombi}, {de Lapparent},
  {Desjacques}, {Gavazzi}, {Kassin}, {Kimm}, {McCracken}, {Milliard},
  {Peirani}, {Prunet}, {Rouberol}, {Silk}, {Slyz}, {Sousbie}, {Teyssier},
  {Tresse}, {Treyer}, {Vibert}, \& {Volonteri}}]{Dubois14}
{Dubois} Y. {et~al.}, 2014, \mnras, 444, 1453

\bibitem[{{Faucher-Gigu{\`e}re} {et~al}\mbox{.}(2011){Faucher-Gigu{\`e}re},
  {Kere{\v{s}}}, \& {Ma}}]{FaucherGiguere11}
{Faucher-Gigu{\`e}re} C.-A., {Kere{\v{s}}} D., {Ma} C.-P., 2011, \mnras, 417,
  2982

\bibitem[{{Finlator} \& {Dav{\'e}}(2008)}]{Finlator08}
{Finlator} K., {Dav{\'e}} R., 2008, \mnras, 385, 2181

\bibitem[{{Guo} {et~al}\mbox{.}(2011){Guo}, {White}, {Boylan-Kolchin}, {De
  Lucia}, {Kauffmann}, {Lemson}, {Li}, {Springel}, \& {Weinmann}}]{Guo11}
{Guo} Q. {et~al.}, 2011, \mnras, 413, 101

\bibitem[{{Haas} {et~al}\mbox{.}(2013){Haas}, {Schaye}, {Booth}, {Dalla
  Vecchia}, {Springel}, {Theuns}, \& {Wiersma}}]{Haas13}
{Haas} M.~R., {Schaye} J., {Booth} C.~M., {Dalla Vecchia} C., {Springel} V.,
  {Theuns} T., {Wiersma} R. P.~C., 2013, \mnras, 435, 2955

\bibitem[{{Hirschmann} {et~al}\mbox{.}(2014){Hirschmann}, {Dolag}, {Saro},
  {Bachmann}, {Borgani}, \& {Burkert}}]{Hirschmann14b}
{Hirschmann} M., {Dolag} K., {Saro} A., {Bachmann} L., {Borgani} S., {Burkert}
  A., 2014, \mnras, 442, 2304

\bibitem[{{Ilbert} {et~al}\mbox{.}(2010){Ilbert}, {Salvato}, {Le Floc'h},
  {Aussel}, {Capak}, {McCracken}, {Mobasher}, {Kartaltepe}, {Scoville},
  {Sanders}, {Arnouts}, {Bundy}, {Cassata}, {Kneib}, {Koekemoer}, {Le
  F{\`e}vre}, {Lilly}, {Surace}, {Taniguchi}, {Tasca}, {Thompson}, {Tresse},
  {Zamojski}, {Zamorani}, \& {Zucca}}]{Ilbert10}
{Ilbert} O. {et~al.}, 2010, \apj, 709, 644

\bibitem[{{Kravtsov} {et~al}\mbox{.}(2018){Kravtsov}, {Vikhlinin}, \&
  {Meshcheryakov}}]{Kravtsov18}
{Kravtsov} A.~V., {Vikhlinin} A.~A., {Meshcheryakov} A.~V., 2018, Astronomy
  Letters, 44, 8

\bibitem[{{Lacey} {et~al}\mbox{.}(2016){Lacey}, {Baugh}, {Frenk}, {Benson},
  {Bower}, {Cole}, {Gonzalez-Perez}, {Helly}, {Lagos}, \& {Mitchell}}]{Lacey16}
{Lacey} C.~G. {et~al.}, 2016, \mnras, 462, 3854

\bibitem[{{Lagos} {et~al}\mbox{.}(2018){Lagos}, {Tobar}, {Robotham},
  {Obreschkow}, {Mitchell}, {Power}, \& {Elahi}}]{Lagos18}
{Lagos} C. d.~P., {Tobar} R.~J., {Robotham} A. S.~G., {Obreschkow} D.,
  {Mitchell} P.~D., {Power} C., {Elahi} P.~J., 2018, \mnras, 481, 3573

\bibitem[{{Larson}(1972)}]{Larson72}
{Larson} R.~B., 1972, Nature Physical Science, 236, 7

\bibitem[{{Larson}(1974)}]{Larson74}
{Larson} R.~B., 1974, \mnras, 169, 229

\bibitem[{{Lilly} {et~al}\mbox{.}(2013){Lilly}, {Carollo}, {Pipino}, {Renzini},
  \& {Peng}}]{Lilly13}
{Lilly} S.~J., {Carollo} C.~M., {Pipino} A., {Renzini} A., {Peng} Y., 2013,
  \apj, 772, 119

\bibitem[{{Lu} {et~al}\mbox{.}(2015){Lu}, {Mo}, {Lu}, {Katz}, {Weinberg}, {van
  den Bosch}, \& {Yang}}]{Lu15b}
{Lu} Z., {Mo} H.~J., {Lu} Y., {Katz} N., {Weinberg} M.~D., {van den Bosch}
  F.~C., {Yang} X., 2015, \mnras, 450, 1604

\bibitem[{{McAlpine} {et~al}\mbox{.}(2016){McAlpine}, {Helly}, {Schaller},
  {Trayford}, {Qu}, {Furlong}, {Bower}, {Crain}, {Schaye}, {Theuns}, {Dalla
  Vecchia}, {Frenk}, {McCarthy}, {Jenkins}, {Rosas-Guevara}, {White}, {Baes},
  {Camps}, \& {Lemson}}]{McAlpine16}
{McAlpine} S. {et~al.}, 2016, Astronomy and Computing, 15, 72

\bibitem[{{Mitchell} {et~al}\mbox{.}(2016){Mitchell}, {Lacey}, {Baugh}, \&
  {Cole}}]{Mitchell16}
{Mitchell} P.~D., {Lacey} C.~G., {Baugh} C.~M., {Cole} S., 2016, \mnras, 456,
  1459

\bibitem[{{Mitchell} {et~al}\mbox{.}(2014){Mitchell}, {Lacey}, {Cole}, \&
  {Baugh}}]{Mitchell14}
{Mitchell} P.~D., {Lacey} C.~G., {Cole} S., {Baugh} C.~M., 2014, \mnras, 444,
  2637

\bibitem[{{Mitchell} {et~al}\mbox{.}(2020{\natexlab{a}}){Mitchell}, {Schaye},
  \& {Bower}}]{Mitchell20b}
{Mitchell} P.~D., {Schaye} J., {Bower} R.~G., 2020{\natexlab{a}}, \mnras, 497,
  4495

\bibitem[{{Mitchell} {et~al}\mbox{.}(2020{\natexlab{b}}){Mitchell}, {Schaye},
  {Bower}, \& {Crain}}]{Mitchell20a}
{Mitchell} P.~D., {Schaye} J., {Bower} R.~G., {Crain} R.~A.,
  2020{\natexlab{b}}, \mnras, 494, 3971

\bibitem[{{Mitra} {et~al}\mbox{.}(2015){Mitra}, {Dav{\'e}}, \&
  {Finlator}}]{Mitra15}
{Mitra} S., {Dav{\'e}} R., {Finlator} K., 2015, \mnras, 452, 1184

\bibitem[{{Moster} {et~al}\mbox{.}(2018){Moster}, {Naab}, \&
  {White}}]{Moster18}
{Moster} B.~P., {Naab} T., {White} S. D.~M., 2018, \mnras, 477, 1822

\bibitem[{{Moster} {et~al}\mbox{.}(2010){Moster}, {Somerville}, {Maulbetsch},
  {van den Bosch}, {Macci{\`o}}, {Naab}, \& {Oser}}]{Moster10}
{Moster} B.~P., {Somerville} R.~S., {Maulbetsch} C., {van den Bosch} F.~C.,
  {Macci{\`o}} A.~V., {Naab} T., {Oser} L., 2010, \apj, 710, 903

\bibitem[{{Muzzin} {et~al}\mbox{.}(2013){Muzzin}, {Marchesini}, {Stefanon},
  {Franx}, {Milvang-Jensen}, {Dunlop}, {Fynbo}, {Brammer}, {Labb{\'e}}, \& {van
  Dokkum}}]{Muzzin13}
{Muzzin} A. {et~al.}, 2013, \apjs, 206, 8

\bibitem[{{Neistein} {et~al}\mbox{.}(2012){Neistein}, {Khochfar}, {Dalla
  Vecchia}, \& {Schaye}}]{Neistein12}
{Neistein} E., {Khochfar} S., {Dalla Vecchia} C., {Schaye} J., 2012, \mnras,
  421, 3579

\bibitem[{{Nelson} {et~al}\mbox{.}(2015){Nelson}, {Genel}, {Vogelsberger},
  {Springel}, {Sijacki}, {Torrey}, \& {Hernquist}}]{Nelson15}
{Nelson} D., {Genel} S., {Vogelsberger} M., {Springel} V., {Sijacki} D.,
  {Torrey} P., {Hernquist} L., 2015, \mnras, 448, 59

\bibitem[{{Oppenheimer} \& {Dav{\'e}}(2008)}]{Oppenheimer08}
{Oppenheimer} B.~D., {Dav{\'e}} R., 2008, \mnras, 387, 577

\bibitem[{{Oppenheimer} {et~al}\mbox{.}(2010){Oppenheimer}, {Dav{\'e}},
  {Kere{\v s}}, {Fardal}, {Katz}, {Kollmeier}, \& {Weinberg}}]{Oppenheimer10}
{Oppenheimer} B.~D., {Dav{\'e}} R., {Kere{\v s}} D., {Fardal} M., {Katz} N.,
  {Kollmeier} J.~A., {Weinberg} D.~H., 2010, \mnras, 406, 2325

\bibitem[{{Pandya} {et~al}\mbox{.}(2021){Pandya}, {Fielding},
  {Angl{\'e}s-Alc{\'a}zar}, {Somerville}, {Bryan}, {Hayward}, {Stern}, {Kim},
  {Quataert}, {Forbes}, {Faucher-Gigu{\`e}re}, {Feldmann}, {Hafen}, {Hopkins},
  {Kere{\v{s}}}, {Murray}, \& {Wetzel}}]{Pandya21}
{Pandya} V. {et~al.}, 2021, \mnras, 508, 2979

\bibitem[{{Pandya} {et~al}\mbox{.}(2020){Pandya}, {Somerville},
  {Angl{\'e}s-Alc{\'a}zar}, {Hayward}, {Bryan}, {Fielding}, {Forbes},
  {Burkhart}, {Genel}, {Hernquist}, {Kim}, {Tonnesen}, \&
  {Starkenburg}}]{Pandya20}
{Pandya} V. {et~al.}, 2020, arXiv e-prints, arXiv:2006.16317

\bibitem[{{Peacock} \& {Smith}(2000)}]{Peacock00}
{Peacock} J.~A., {Smith} R.~E., 2000, \mnras, 318, 1144

\bibitem[{{Peeples} {et~al}\mbox{.}(2014){Peeples}, {Werk}, {Tumlinson},
  {Oppenheimer}, {Prochaska}, {Katz}, \& {Weinberg}}]{Peeples14}
{Peeples} M.~S., {Werk} J.~K., {Tumlinson} J., {Oppenheimer} B.~D., {Prochaska}
  J.~X., {Katz} N., {Weinberg} D.~H., 2014, \apj, 786, 54

\bibitem[{{Pillepich} {et~al}\mbox{.}(2018){Pillepich}, {Springel}, {Nelson},
  {Genel}, {Naiman}, {Pakmor}, {Hernquist}, {Torrey}, {Vogelsberger},
  {Weinberger}, \& {Marinacci}}]{Pillepich18}
{Pillepich} A. {et~al.}, 2018, \mnras, 473, 4077

\bibitem[{{Planck Collaboration} {et~al}\mbox{.}(2014){Planck Collaboration},
  {Ade}, {Aghanim}, {Armitage-Caplan}, {Arnaud}, {Ashdown}, {Atrio-Barandela},
  {Aumont}, {Baccigalupi}, {Banday}, \& et~al.}]{Planck14}
{Planck Collaboration} {et~al.}, 2014, \aap, 571, A16

\bibitem[{{Qu} {et~al}\mbox{.}(2017){Qu}, {Helly}, {Bower}, {Theuns}, {Crain},
  {Frenk}, {Furlong}, {McAlpine}, {Schaller}, {Schaye}, \& {White}}]{Qu17}
{Qu} Y. {et~al.}, 2017, \mnras, 464, 1659

\bibitem[{{Rees} \& {Ostriker}(1977)}]{Rees77}
{Rees} M.~J., {Ostriker} J.~P., 1977, \mnras, 179, 541

\bibitem[{{Rodr{\'\i}guez-Puebla} {et~al}\mbox{.}(2017){Rodr{\'\i}guez-Puebla},
  {Primack}, {Avila-Reese}, \& {Faber}}]{RodriguezPuebla17}
{Rodr{\'\i}guez-Puebla} A., {Primack} J.~R., {Avila-Reese} V., {Faber} S.~M.,
  2017, \mnras, 470, 651

\bibitem[{{Schaye}(2004)}]{Schaye04}
{Schaye} J., 2004, \apj, 609, 667

\bibitem[{{Schaye} {et~al}\mbox{.}(2015){Schaye}, {Crain}, {Bower}, {Furlong},
  {Schaller}, {Theuns}, {Dalla Vecchia}, {Frenk}, {McCarthy}, {Helly},
  {Jenkins}, {Rosas-Guevara}, {White}, {Baes}, {Booth}, {Camps}, {Navarro},
  {Qu}, {Rahmati}, {Sawala}, {Thomas}, \& {Trayford}}]{Schaye15}
{Schaye} J. {et~al.}, 2015, \mnras, 446, 521

\bibitem[{{Schaye} \& {Dalla Vecchia}(2008)}]{Schaye08}
{Schaye} J., {Dalla Vecchia} C., 2008, \mnras, 383, 1210

\bibitem[{{Schaye} {et~al}\mbox{.}(2010){Schaye}, {Dalla Vecchia}, {Booth},
  {Wiersma}, {Theuns}, {Haas}, {Bertone}, {Duffy}, {McCarthy}, \& {van de
  Voort}}]{Schaye10}
{Schaye} J. {et~al.}, 2010, \mnras, 402, 1536

\bibitem[{{Sharma} \& {Theuns}(2020)}]{Sharma20}
{Sharma} M., {Theuns} T., 2020, \mnras, 492, 2418

\bibitem[{{Silk} \& {Rees}(1998)}]{Silk98}
{Silk} J., {Rees} M.~J., 1998, \aap, 331, L1

\bibitem[{{Somerville} {et~al}\mbox{.}(2012){Somerville}, {Gilmore}, {Primack},
  \& {Dom{\'{\i}}nguez}}]{Somerville12}
{Somerville} R.~S., {Gilmore} R.~C., {Primack} J.~R., {Dom{\'{\i}}nguez} A.,
  2012, \mnras, 423, 1992

\bibitem[{{Somerville} {et~al}\mbox{.}(2008){Somerville}, {Hopkins}, {Cox},
  {Robertson}, \& {Hernquist}}]{Somerville08}
{Somerville} R.~S., {Hopkins} P.~F., {Cox} T.~J., {Robertson} B.~E.,
  {Hernquist} L., 2008, \mnras, 391, 481

\bibitem[{{Springel}(2005)}]{Springel05b}
{Springel} V., 2005, \mnras, 364, 1105

\bibitem[{{Springel} {et~al}\mbox{.}(2001){Springel}, {White}, {Tormen}, \&
  {Kauffmann}}]{Springel01}
{Springel} V., {White} S.~D.~M., {Tormen} G., {Kauffmann} G., 2001, \mnras,
  328, 726

\bibitem[{{Tabor} \& {Binney}(1993)}]{Tabor93}
{Tabor} G., {Binney} J., 1993, \mnras, 263, 323

\bibitem[{{{\"U}bler} {et~al}\mbox{.}(2014){{\"U}bler}, {Naab}, {Oser},
  {Aumer}, {Sales}, \& {White}}]{Ubler14}
{{\"U}bler} H., {Naab} T., {Oser} L., {Aumer} M., {Sales} L.~V., {White} S.
  D.~M., 2014, \mnras, 443, 2092

\bibitem[{{Vale} \& {Ostriker}(2004)}]{Vale04}
{Vale} A., {Ostriker} J.~P., 2004, \mnras, 353, 189

\bibitem[{{van de Voort}(2017)}]{VanDeVoort17b}
{van de Voort} F., 2017, Astrophysics and Space Science Library, Vol. 430, {The
  Effect of Galactic Feedback on Gas Accretion and Wind Recycling}, {Fox} A.,
  {Dav{\'e}} R., eds., p. 301

\bibitem[{{van de Voort} {et~al}\mbox{.}(2011){van de Voort}, {Schaye},
  {Booth}, {Haas}, \& {Dalla Vecchia}}]{VanDeVoort11}
{van de Voort} F., {Schaye} J., {Booth} C.~M., {Haas} M.~R., {Dalla Vecchia}
  C., 2011, \mnras, 414, 2458

\bibitem[{{Vogelsberger} {et~al}\mbox{.}(2014){Vogelsberger}, {Genel},
  {Springel}, {Torrey}, {Sijacki}, {Xu}, {Snyder}, {Nelson}, \&
  {Hernquist}}]{Vogelsberger14}
{Vogelsberger} M. {et~al.}, 2014, \mnras, 444, 1518

\bibitem[{{Wang} {et~al}\mbox{.}(2013){Wang}, {Farrah}, {Oliver}, {Amblard},
  {B{\'e}thermin}, {Bock}, {Conley}, {Cooray}, {Halpern}, {Heinis}, {Ibar},
  {Ilbert}, {Ivison}, {Marsden}, {Roseboom}, {Rowan-Robinson}, {Schulz},
  {Smith}, {Viero}, \& {Zemcov}}]{Wang13}
{Wang} L. {et~al.}, 2013, \mnras, 431, 648

\bibitem[{{White} \& {Frenk}(1991)}]{White91}
{White} S.~D.~M., {Frenk} C.~S., 1991, \apj, 379, 52

\bibitem[{{White} \& {Rees}(1978)}]{White78}
{White} S.~D.~M., {Rees} M.~J., 1978, \mnras, 183, 341

\bibitem[{{Wiersma} {et~al}\mbox{.}(2009){Wiersma}, {Schaye}, {Theuns}, {Dalla
  Vecchia}, \& {Tornatore}}]{Wiersma09}
{Wiersma} R.~P.~C., {Schaye} J., {Theuns} T., {Dalla Vecchia} C., {Tornatore}
  L., 2009, \mnras, 399, 574

\bibitem[{{Wright} {et~al}\mbox{.}(2020){Wright}, {Lagos}, {Power}, \&
  {Mitchell}}]{Wright20}
{Wright} R.~J., {Lagos} C. d.~P., {Power} C., {Mitchell} P.~D., 2020, \mnras

\bibitem[{{Yang} {et~al}\mbox{.}(2012){Yang}, {Mo}, {van den Bosch}, {Zhang},
  \& {Han}}]{Yang12}
{Yang} X., {Mo} H.~J., {van den Bosch} F.~C., {Zhang} Y., {Han} J., 2012, \apj,
  752, 41

\end{thebibliography}

\appendix

\section{}
\label{ap_bias_plots}

\begin{figure}
\includegraphics[width=20pc]{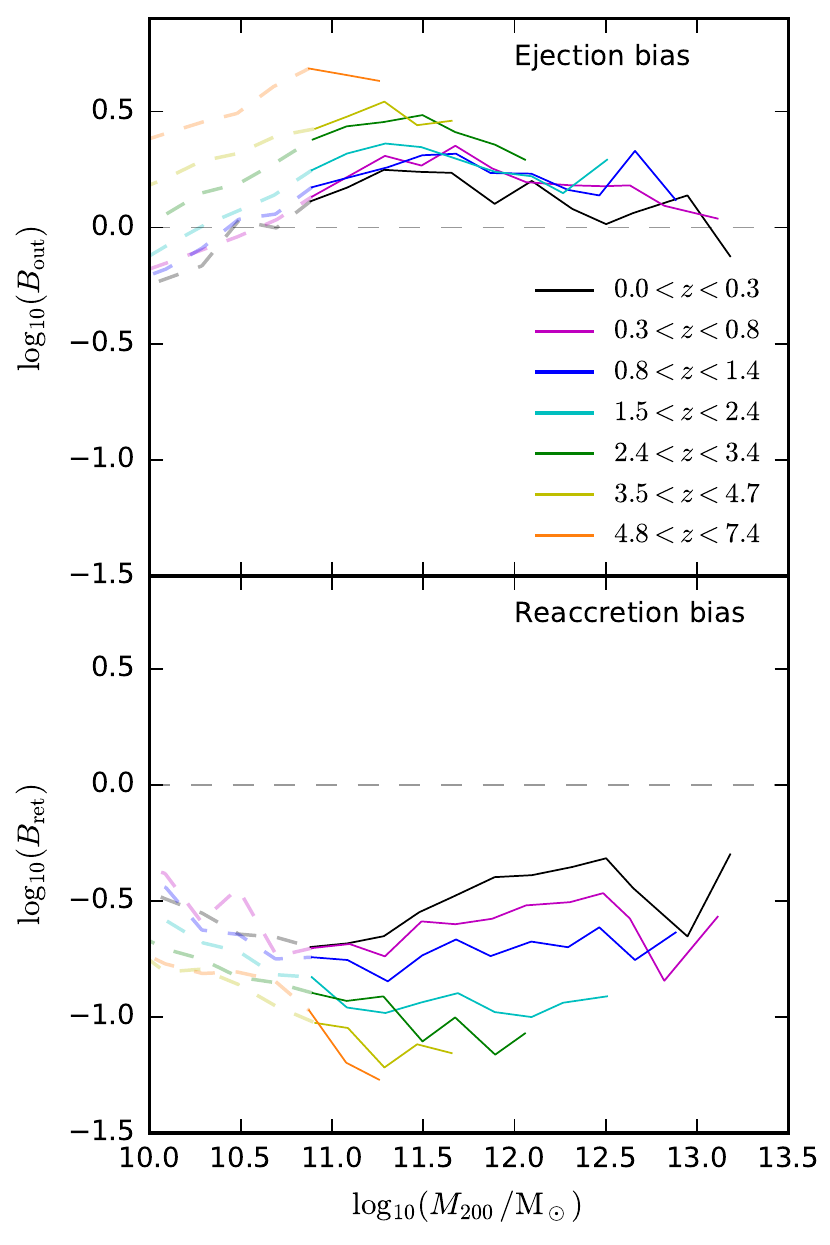}
\caption{
\textit{Top:} the dependence of the bias parameter $B_{\mathrm{out}}$ 
on halo mass and redshift \protect (defined in Eqn~\ref{bias_out_eqn}).
$B_{\mathrm{out}}$ expresses whether
any gas that is ejected outside of haloes is biased with respect
to whether that gas was previously part of the ISM. 
$B_{\mathrm{out}} >1$ implies that former-ISM circum-galactic
gas is more likely to be ejected from the halo than gas that
has not been processed through the ISM.
\textit{Bottom:} the corresponding halo mass and redshift
dependence of the second bias parameter $B_{\mathrm{ret}}$
\protect (see Eqn~\ref{bias_ret_eqn}).
$B_{\mathrm{ret}}$ expresses whether any gas that is re-accreted
onto haloes is biased with respect to whether that gas
was previously part of the ISM.
}
\label{bias_plot}
\end{figure}

As briefly mentioned at the end of Section~\ref{n12_desc_sec},
the computation of the quantity $F_{\mathrm{CGM}}^{\mathrm{pr}}$ that appears
in Eqn.~\ref{ODE_MATRIX} requires some additional steps. 
As a reminder, $F_{\mathrm{CGM}}^{\mathrm{pr}}$ is the mass fraction
of the CGM that has \emph{not} been ejected from the ISM of a progenitor
galaxy in the past. In practice, our model internally computes the
mass within two separate CGM mass reservoirs: splitting between gas that
has, and has not, been part of the ISM previously (expressing this in terms of
$F_{\mathrm{CGM}}^{\mathrm{pr}}$ then conveniently reduces the associated equations
down to the form seen in Eqn.~\ref{ODE_MATRIX}).

At the same time, we prefer to define outflows at the scale
of the virial radius with a single mass loading factor, $\eta^{\mathrm{halo}}$.
In isolation, $\eta^{\mathrm{halo}}$ does not determine how much
of the ejected gas came from the ISM, and how much did not.
Similarly, we define halo-scale gas recycling with a single coefficient
$G_{\mathrm{ret}}^{\mathrm{halo}}$, which again does not specify how
much of the returning gas was part of the ISM in the past.

These two terms are therefore supplemented with two additional 
terms, labelled $B_{\mathrm{out}}$ and $B_{\mathrm{ret}}$ which
do specify the fraction of halo-scale ejected/returning gas that was part
of the ISM in the past. Note that neither $B_{\mathrm{out}}$ nor
$B_{\mathrm{ret}}$ affect (directly) the total mass in the CGM ($M_{\mathrm{CGM}}$), and
so do not appear in Eqn.~\ref{ODE_MATRIX}.

The two terms are defined as

\begin{equation}
B_{\mathrm{out}} \equiv \frac{F_{\mathrm{CGM,out}}^{\mathrm{gal}}}{F_{\mathrm{CGM}}^{\mathrm{gal}}},
\label{bias_out_eqn}
\end{equation}

\noindent and

\begin{equation}
B_{\mathrm{ret}} \equiv \frac{F_{\mathrm{CGM,ret}}^{\mathrm{gal}}}{F_{\mathrm{ej,halo}}^{\mathrm{gal}}},
\label{bias_ret_eqn}
\end{equation}

\noindent where $F_{\mathrm{CGM}}^{\mathrm{gal}}$ ($=1 - F_{\mathrm{CGM}}^{\mathrm{pr}}$)
is the mass fraction of the CGM that was previously ejected from the ISM of a progenitor galaxy, and
$F_{\mathrm{CGM,out}}^{\mathrm{gal}}$
is the fraction of the halo-scale mass outflow rate that was previously ejected from the ISM
of a progenitor galaxy.
Similarly, $F_{\mathrm{ej,halo}}^{\mathrm{gal}}$
is the mass fraction of the ejected gas reservoir (outside $R_{200}$) that was also ejected from the ISM of a progenitor
galaxy, and
$F_{\mathrm{CGM,ret}}^{\mathrm{gal}}$
is the fraction of the halo-scale recycled mass inflow rate that was previously
ejected from the ISM of a progenitor galaxy.

As with the other coefficients introduced
in Eqn.~\ref{ODE_MATRIX}, we compute $B_{\mathrm{out}}$
and $B_{\mathrm{ret}}$ from an \eagle simulation,
averaging as a function of halo mass and redshift.
When the values of $B_{\mathrm{out}}$ and $B_{\mathrm{ret}}$ are 
specified, we can then compute the desired quantities 
$F_{\mathrm{CGM,out}}^{\mathrm{gal}}$ and $F_{\mathrm{CGM,ret}}^{\mathrm{gal}}$,
since $F_{\mathrm{CGM}}^{\mathrm{gal}}$ ($=1 - F_{\mathrm{CGM}}^{\mathrm{pr}}$) and $F_{\mathrm{ej,halo}}^{\mathrm{gal}}$
are both quantities that are tracked internally within
the model.

Note that the definitions of $B_{\mathrm{out}}$ and $B_{\mathrm{ret}}$ are constructed 
such that (for example) if the ISM-processed
mass fraction of the CGM ($F_{\mathrm{CGM}}^{\mathrm{gal}}$) increases with
respect to the values recorded in the \eagle simulation, then for fixed
$B_{\mathrm{out}}$ the model responds accordingly by increasing the ISM-processed
fraction of the halo-scale outflow rate.

With the two terms $B_{\mathrm{out}}$ and $B_{\mathrm{ret}}$ specified,
we can then track separately the fraction of mass in the CGM
($M_{\mathrm{CGM}}$) and in the ejected gas reservoir outside $R_{200}$ ($M_{\mathrm{ej}}^{\mathrm{halo}}$)
that was part of the ISM in the past.
Neither term strongly affects our results, and we maintain both at their
fiducial values throughout our analysis. 
As the definition of these parameters was only
introduced relatively late in the development of this
project, we measure $B_{\mathrm{out}}$ and $B_{\mathrm{ret}}$
from a smaller version of the main reference \eagle
simulation, using instead a $(25 \, \mathrm{Mpc})^3$
volume. We extrapolate to higher halo masses simply
by holding the values of $B_{\mathrm{out}}$ and $B_{\mathrm{ret}}$
fixed to the end points of the measured range.

These dependencies are shown in Fig.~\ref{bias_plot}.
Circum-galactic gas that was part of the ISM in the past
is generally more likely to be ejected out of the halo than
circum-galactic gas that was not formerly part of the ISM
(i.e., $B_{\mathrm{out}} > 1$).
Conversely, ejected gas residing outside $R_{200}$ is
more likely to be re-accreted if it was not part of the ISM
in the past (i.e., $B_{\mathrm{ret}} < 1$). $B_{\mathrm{out}}$ and $B_{\mathrm{ret}}$
do not have a large effect on our results for galaxy
stellar masses, and we hold them at their fiducial values
throughout our analysis.

\section{}
\label{ap_dm_first}

\begin{figure*}
\begin{center}
\includegraphics[width=40pc]{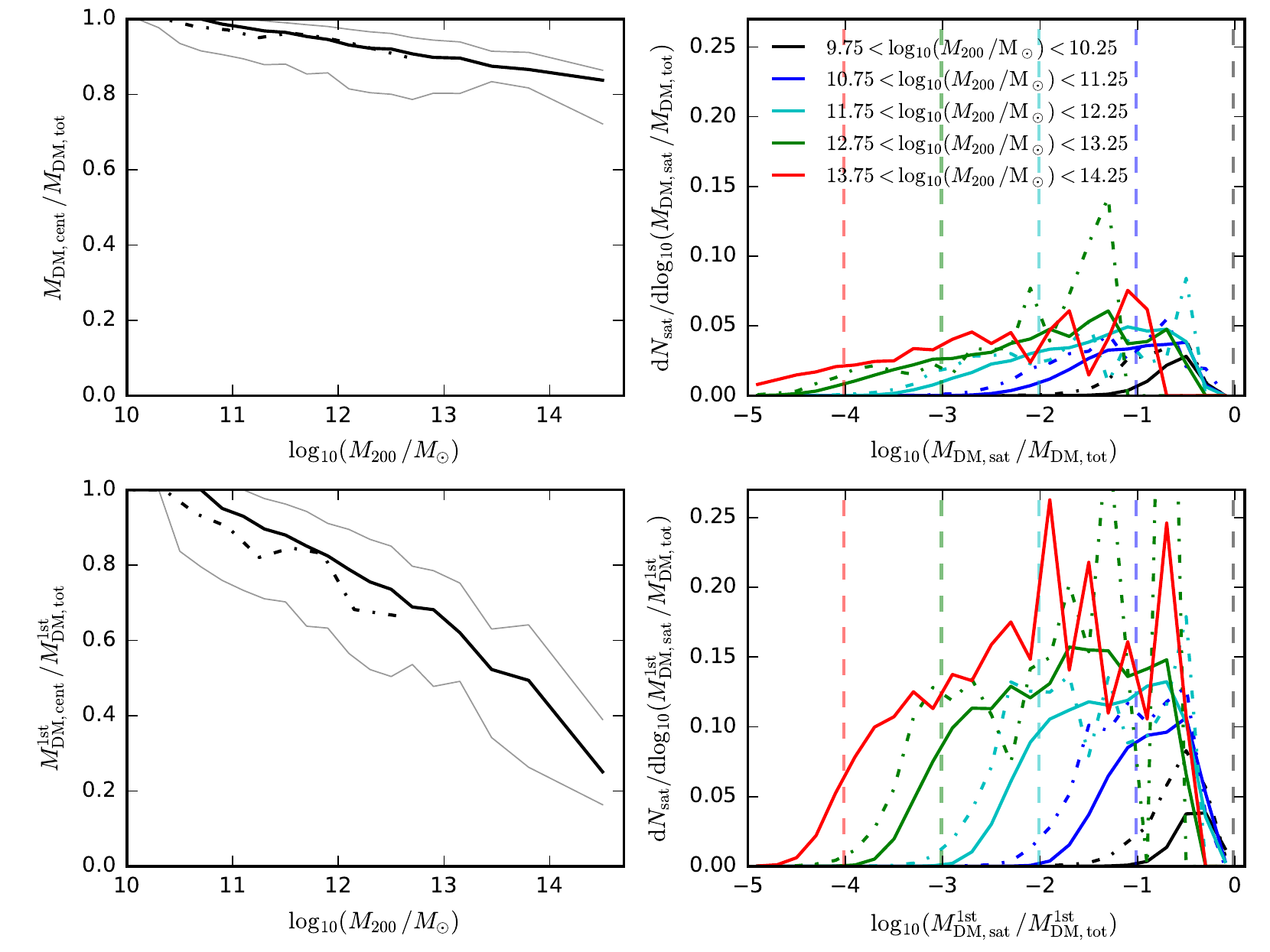}
\caption{\textit{Left panels:} 
the fraction of mass in central subhaloes at $z=0$, compared to the total halo mass including satellite subhaloes,
plotted as a function of halo mass. 
The top panel shows the mass fraction for the instantaneous dark matter mass as 
identified by \subfind ($M_{\mathrm{DM}}$). 
The bottom panel shows the cumulative mass of dark matter particles that have been accreted for
the first time onto a given subhalo and its progenitors ($M_{\mathrm{DM}}^{\mathrm{1st}}$).
Solid lines show the $16$, $50$, and $84^{\mathrm{th}}$ percentiles for the Reference \eagle simulation.
Dash-dotted lines show the median relation for a smaller-volume simulation with $8$ times higher particle resolution (Recal).
For all halo masses, the instantaneous dark matter mass is always dominated by the central subhalo;
satellite subhaloes generally lose most of their mass to tidal stripping.
Conversely, satellite subhaloes make up the majority of $M_{\mathrm{DM}}^{\mathrm{1st}}$ for
$M_{200} > 10^{13.5} \, \mathrm{M_\odot}$, since $M_{\mathrm{DM}}^{\mathrm{1st}}$ is not affected by stripping.
$M_{\mathrm{DM}}^{\mathrm{1st}}$ is expected to be more closely connected to the stellar mass
of satellite galaxies than $M_{\mathrm{DM}}$, as stars are more tightly bound on average
than dark matter particles.
\textit{Right panels:}
the differential distribution of satellite subhalo masses for $M_{\mathrm{DM}}$ (top) 
and $M_{\mathrm{DM}}^{\mathrm{1st}}$ (bottom).
Different line colours indicate different bins of mass of the corresponding host halo.
Dashed vertical lines indicate (for the Reference \eagle simulation) the scale for
which the satellite subhalo mass is less than the mass of $100$ dark matter particles.
Below this scale, satellite subhaloes are not well resolved, and so the simulation becomes
incomplete.
Comparing the Reference simulation (solid lines) to Recal (dash-dotted lines), it is evident that
the simulation is not converged in terms of the total mass in satellites, especially for 
$M_{\mathrm{DM}}^{\mathrm{1st}}$.
}
\label{mfrac_sub}
\end{center}
\end{figure*}

In the lower-right panel of Fig.~\ref{m12_fix_fig} of Section~\ref{shm_shape_subsec},
we showed that if all explicit halo mass dependence is removed from the N12 model
(by fixing each model term to its value at $M_{200} = 10^{12} \, \mathrm{M_\odot}$),
the SHM relation becomes nearly flat (i.e. $M_\star/M_{200} \approx$ constant), but 
still drops significantly for $M_{200} > 10^{13} \, \mathrm{M_\odot}$.
The reason for this drop at high halo masses is demonstrated in Fig.~\ref{mfrac_sub},
which shows the fraction of mass in central subhaloes (as opposed to satellites) in
the left-side panels, and the differential distribution of satellite subhalo masses
in the right-side panels. 

We show two distinct definitions of subhalo mass: the
instantaneous dark matter subhalo mass ($M_{\mathrm{DM}}$, top panels), and the 
cumulative mass of dark matter accreted onto subhaloes ($M_{\mathrm{DM}}^{\mathrm{1st}}$).
The latter quantity is computed by summing over all first-time dark matter accretion
onto all progenitors of the final subhalo, and is directly connected to the source
term in our implementation of the N12 model (see Eqn.~\ref{ODE_MATRIX}). The main distinction between $M_{\mathrm{DM}}$
and $M_{\mathrm{DM}}^{\mathrm{1st}}$ is that only $M_{\mathrm{DM}}$ is affected by processes
that remove dark matter from a subhalo, which for satellites is dominated by tidal
stripping processes as the satellite orbits within the tidal field exerted by the host.
On average, stars are significantly more bound than dark matter particles to a subhalo,
and are accordingly much less effected by tidal stripping. As such, we would expect
that satellite stellar masses would be more closely related to $M_{\mathrm{DM}}^{\mathrm{1st}}$
than to $M_{\mathrm{DM}}$.

The upper-left panel of Fig.~\ref{mfrac_sub} shows that the central subhalo
always dominates the instantaneous mass of dark matter haloes within \eagle. 
The mass fraction within the central subhalo does decrease slightly with
increasing halo mass, dropping from $\approx 100 \, \%$ for 
$M_{200} = 10^{11} \, \mathrm{M_\odot}$ to $\approx 80 \, \%$ for 
$M_{200} = 10^{14.5} \, \mathrm{M_\odot}$. This is expected, partly because higher-mass
haloes are less relaxed because they collapsed more recently, which implies
more substructure (leading also to a weak, negative halo concentration-mass
relation), and partly since \eagle resolves a
larger number of substructures in high-mass haloes relative to low-mass
haloes. 

For $M_{\mathrm{DM}}^{\mathrm{1st}}$ (lower-left panel), satellite subhaloes make a significantly
larger contribution to the total mass fraction, and actually comprise the majority
of the mass for $M_{200} > 10^{13.5}$. The strong dependence of the $M_{\mathrm{DM}}^{\mathrm{1st}}$ 
mass fraction on halo mass is again expected to be partly a resolution effect,
which is stronger in this case. If we compare a simulation at
the standard \eagle resolution 
(solid lines, with numerical particle mass of $m_{\mathrm{DM}} = 9.7 \times 10^7 \, \mathrm{M_\odot}$)
to an eight-times higher mass resolution simulation (dash-dotted lines) we see indeed
that the fraction of $M_{\mathrm{DM}}^{\mathrm{1st}}$ is lower at higher resolution.
The right panels of Fig.~\ref{mfrac_sub} show the differential distribution of
satellite masses for the two simulations, demonstrating that the higher-resolution
simulation resolves more of the low-mass substructure, as expected.

We emphasise that changes in satellite contributions to $M_{\mathrm{DM}}^{\mathrm{1st}}$
with changing resolution are not ordinarily a large concern for galaxy stellar masses. This is because
low-mass satellites (with $M_{\mathrm{DM}}^{\mathrm{1st}} \ll 10^{12} \, \mathrm{M_\odot}$)
are far less efficient at forming stars than satellites with 
$M_{\mathrm{DM}}^{\mathrm{1st}} \sim 10^{12} \, \mathrm{M_\odot}$, which are resolved
in simulations like \eagle. For the unrealistic model variation shown in Fig.~\ref{m12_fix_fig}
(yellow line), low-mass satellites are just as efficient at forming stars as high-mass satellites.
In this case therefore, not resolving a population of low-mass satellites will be an issue,
contributing to the decrease in the SHM relation for $M_{200} > 10^{13} \, \mathrm{M_\odot}$,
since these haloes do resolve a larger fraction of the (now-relevant) low-mass substructures.

\label{lastpage}
\end{document}